\documentclass[aps,twocolumn,superscriptrddress,10pt,showpacs]{revtex4-1}
\pdfoutput=1

\usepackage{graphicx}
\usepackage{amsmath,amssymb}
\usepackage{hyperref}
\hypersetup{colorlinks=true,
						linkcolor=red,
						anchorcolor=blue,
						citecolor=blue,
						filecolor=blue,
						menucolor=blue,
						urlcolor=blue
}

\usepackage{datetime}
\usepackage{color}
\usepackage[english]{babel}
\usepackage{xspace}
\usepackage{textcomp}
\newcommand{\bra}[1]{\langle #1|}
\newcommand{\ket}[1]{\ensuremath{|#1\rangle}\xspace}

\newcommand{\pdo}{\ensuremath{p_{\mathrm{DO}}}\xspace}
\newcommand{\pdomax}{\ensuremath{p_{\mathrm{DO}}^{\mathrm{(max)}}}\xspace}
\newcommand{\psmax}{\ensuremath{p_{\mathrm{s}}^{\mathrm{(max)}}}\xspace}
\newcommand{\ps}{\ensuremath{p_{\mathrm s}}\xspace}
\newcommand{\pt}{\ensuremath{p_{\mathrm t}}\xspace}
\newcommand{\pdor}{\ensuremath{p_{\mathrm{DO}}^{(R)}}\xspace}
\newcommand{\psr}{\ensuremath{p_{\mathrm s}^{(R)}}\xspace}

\newcommand{\pbdo}{\ensuremath{\bar{p}_{\mathrm{DO}}}\xspace}
\newcommand{\pbs}{\ensuremath{\bar{p}_{\mathrm s}}\xspace}

\newcommand{\pbdor}{\ensuremath{\bar{p}_{\mathrm{DO}}^{(R)}}\xspace}
\newcommand{\pbsr}{\ensuremath{\bar{p}_{\mathrm s}^{(R)}}\xspace}

\newcommand{\Uload}{\ensuremath{U_{\mathrm{load}}}\xspace}
\newcommand{\ketstilde}[0]{\ensuremath{|\tilde{\mathrm s}\rangle}\xspace}

\newcommand{\ketptilde}[0]{\ensuremath{|\tilde{\mathrm D}_+\rangle}\xspace}
\newcommand{\ketmtilde}[0]{\ensuremath{|\tilde{\mathrm D}_-\rangle}\xspace}
\begin{document}

\title{Controlling the Floquet state population and observing micromotion in a periodically driven two-body quantum system}

\author{R\'emi Desbuquois, Michael Messer, Frederik G\"org, Kilian Sandholzer, Gregor Jotzu, Tilman Esslinger}
\affiliation{Institute for Quantum Electronics, ETH Zurich, 8093 Zurich, Switzerland}

\date{\today}

\begin{abstract}

Near-resonant periodic driving of quantum systems promises the implementation of a large variety of novel quantum states, though their preparation and measurement remains challenging.
We address these aspects in a model system consisting of interacting fermions in a periodically driven array of double wells created by an optical lattice.
The singlet and triplet fractions and the double occupancy of the Floquet states are measured, and their behavior as a function of the interaction strength is analyzed in the high- and low-frequency regimes. 
We demonstrate full control of the Floquet state population and find suitable ramping protocols and time-scales which adiabatically connect the initial ground state to different targeted Floquet states. 
The micromotion which exactly describes the time evolution of the system within one driving cycle is observed. 
Additionally, we provide an analytic description of the model and compare it to numerical simulations. 
\end{abstract}

\maketitle

Floquet engineering aims to create novel quantum states through periodic driving, by realising effective Hamiltonians beyond the reach of static systems \cite{Goldman2014a, Bukov2015b, Eckardt2017}.
These effective Hamiltonians have been implemented with photons \cite{Rechtsman2013, Sommer2016}, solids \cite{Wang2013b} and ultracold gases in optical lattices \cite{Eckardt2017}.
However, preparing and controlling a specific quantum state in a driven system remains in general a challenge.
This is particularly the case for many interesting schemes which were realised by driving at low frequencies \cite{Rechtsman2013, Jotzu2014} or even close to a characteristic energy scale of the underlying static Hamiltonian.
Indeed, driving near-resonantly to the band structure was used to modify kinetic terms in the Hamiltonian \cite{Chen2011, Parker2013, Williams2013, Aidelsburger2014, Kennedy2015, Weinberg2015, Khamehchi2016, Flaschner2016}, and modulating close to the interaction energy was proposed to engineer novel interaction terms \cite{Mentink2015, Kitamura2016, Coulthard2016, Strater2016, Kitamura2017}. 
For all these schemes, the periodic drive strongly couples the static eigenstates, which makes the full control of the population of the different Floquet states and the analysis of their exact time evolution demanding.

One important aspect lies in the fundamental differences between Floquet-engineered systems and static Hamiltonians.
For example, a periodically driven system is described by a periodic quasi-energy spectrum, and thus has no ground state. 
Its absence raises an important experimental challenge: How to adiabatically connect the ground state of the initial static Hamiltonian to the targeted Floquet eigenstate?
Theory suggests that the population of Floquet states has a non-trivial dependence on the ramp speed and on the exact trajectory which is used in parameter-space \cite{Breuer1989, Breuer1989a, Hone1997, Eckardt2008, Ho2016, Weinberg2016}, particularly in the case of near-resonant driving which leads to the formation of avoided crossings between quasi-energy levels \cite{Grifoni1998}.
In addition to this aspect, we now have to measure observables that are affected by micromotion describing the dynamics of the Floquet system within a driving period.
Whilst this micromotion tends to become negligible for infinite driving frequencies, it alters the states significantly for near-resonant and low-frequency modulation \cite{Greif2011, Rechtsman2013, Goldman2014a, Bukov2014a, Bukov2015b, Goldman2015, Anisimovas2015}.

\begin{figure}[b!]
    \includegraphics{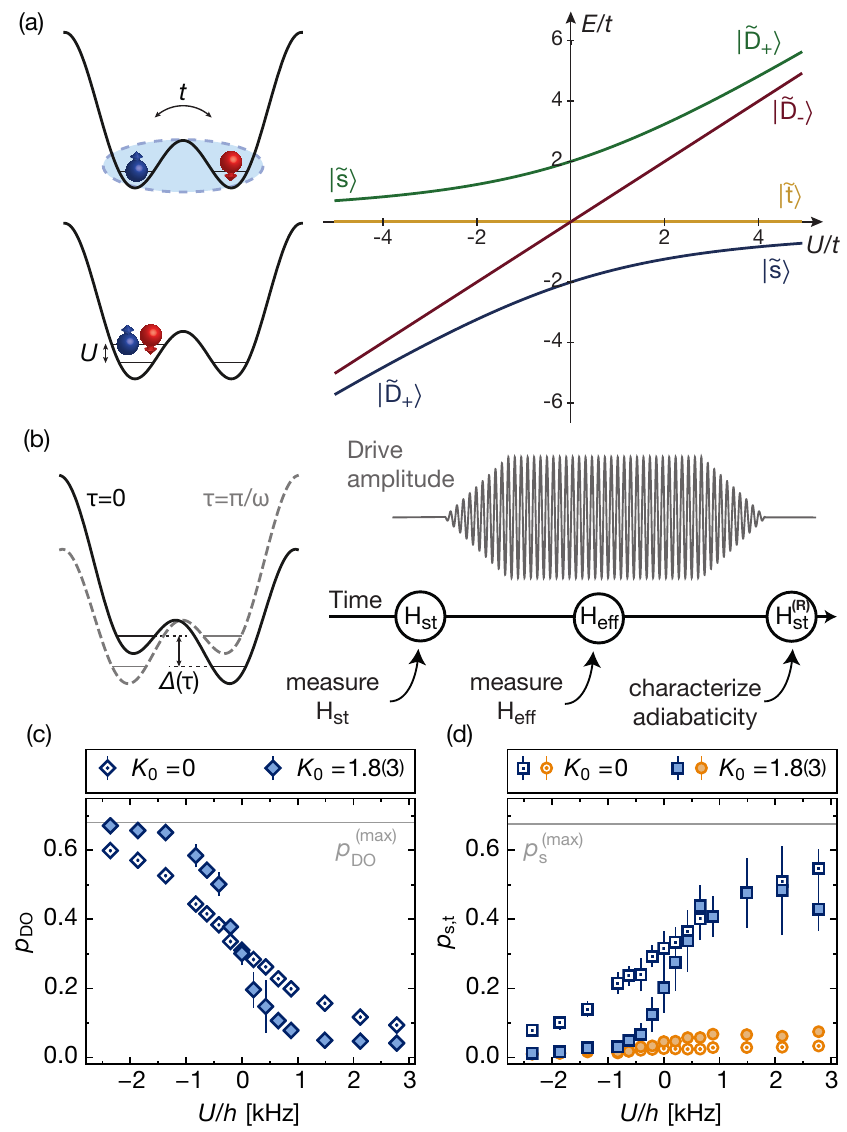}
    \caption{
			Off-resonant lattice modulation of a double well array. 
			(a) Schematic view of a single double well (left) and its spectrum as a function of the on-site interaction $U$ in units of the tunneling $t$ (right). 
			(b) The driven regime is reached while ramping up a sinusoidal modulation of the lattice position which is equivalent to modulating the site offset $\Delta(\tau)$.  
			By quenching the tunneling to zero during the modulation (at the point $H_{\mathrm{eff}}$), we freeze the evolution of the quantum states to measure their population in the effective Floquet Hamiltonian $H_{\mathrm{eff}}$. 
			Reverting the modulation ramp and then subsequently quenching the tunneling allows to determine the adiabaticity of the Floquet engineering process $H_{\mathrm{st}}$, at the point $H_{\mathrm{st}}^{(R)}$.
			(c) Measurement of the double-occupancy fraction \pdo as a function of $U$ of the static Hamiltonian (blue open-dotted points) and the effective Hamiltonian for off-resonant driving at $\omega/2\pi=8 $ kHz at shaking amplitude $K_0=1.8(3)$ (blue filled points).  
			(d) Singlet \ps (blue) and triplet fraction \pt (orange) measured for the static Hamiltonian (open-dotted data points) and the effective Hamiltonian (filled data points).
			The grey horizontal lines indicate the maximal possible fractions of \pdomax and \psmax resulting from the initial preparation of the system. 
			Error bars denote the standard deviation of 5 measurements.}
	\label{fig1}
\end{figure}

In this work, we address the previously mentioned challenges in a tractable way by realizing a periodically driven array of double wells \cite{Kierig2008} occupied by pairs of interacting atoms \cite{Sebby-Strabley2006, Folling2007, Greif2013a, Murmann2014a}, which allows for a full control of the Floquet state population \cite{Grifoni1998}.
The symmetric double well consists of two sites containing two opposite spins, and can be described by a Hubbard model with a tunneling amplitude $t$, and an on-site interaction $U$. 
In Appendix~\ref{ap:analytic} we provide a detailed explanation and derivation of the realized Hamiltonian.
The resulting Hilbert space is spanned by the singlet state $\ket{\mathrm s} = (\ket{\mathord{\uparrow},\downarrow}-\ket{\mathord{\downarrow}, \uparrow})/\sqrt{2}$ and the triplet state $\ket{\mathrm t} = (\ket{\mathord{\uparrow},\downarrow}+\ket{\mathord{\downarrow}, \uparrow})/\sqrt{2}$ where both sites are occupied, and by the states $\ket{\mathrm{D}_\pm} = (\ket{\mathord{\uparrow\downarrow},0}\pm\ket{0,\uparrow\downarrow})/\sqrt{2}$ where one site is doubly occupied.
In this basis, the Hamiltonian is given by 
\begin{equation*}
H=-2\,t\big(\ket{\mathrm s}\bra{\mathrm{D}_+}+\ket{\mathrm{D}_+}\bra{\mathrm s}\big)+U\big(\ket{\mathrm{D}_-}\bra{\mathrm{D}_-}+\ket{\mathrm{D}_+}\bra{\mathrm{D}_+}\big), 
\end{equation*}
and its spectrum is shown in Fig. \ref{fig1}(a).
The ground state smoothly evolves from $\ket{\mathrm{D}_+}$ to $\ket{\mathrm s}$ as the interactions are tuned from strongly attractive to strongly repulsive.
The two components are equally populated for $U=0$, and the width of the crossover region is given by $4t$.
In Fig. \ref{fig1}(a) and in the rest of the paper, the energy levels are labeled by the corresponding state in the large $U$-limit with a tilde.
In our notation, the ground state is thus labeled \ketptilde for negative $U$ and  \ketstilde for positive $U$.
The periodic drive consists in a modulation in time of the potential bias $\Delta$ between the two wells (see Fig. \ref{fig1}(b)), which couples the states $\ket{\mathrm{D}_\pm}$ via 
\begin{equation*}
\hat{V}(\tau)=\Delta(\tau)(\ket{\mathrm{D}_+}\bra{\mathrm{D}_-}+\ket{\mathrm{D}_-}\bra{\mathrm{D}_+}).
\end{equation*}

To realize this system, we begin our experiment with $N_{\mathrm{tot}}= 159(10)\times 10^3$ (15\% systematic error) ultracold fermionic $^{40}$K atoms, harmonically confined, in a balanced two-component spin mixture prepared in the two magnetic sublevels $m_F = -9/2, -7/2$ of the $F = 9/2$ hyperfine manifold \cite{Greif2013a}.
We load the atoms into an array of isolated double wells created with a tunable-geometry optical lattice \cite{Tarruell2012, si}.
After this procedure, 68(3) \% of the double wells contain two opposite spins. 
The tunneling amplitude $t$ between the two wells can be tuned by changing the depth of the lattice while keeping the tunneling amplitude to neighboring dimers at a negligible value of less than $h\times 3$ Hz.
The interaction strength $U$ is then set to the desired value by tuning the s-wave scattering length via a magnetic Feshbach resonance.
To enter the driven regime, we linearly ramp up a sinusoidal modulation of the lattice position with frequency $\omega/2\pi$ along the direction of the dimers \cite{si}, and then maintain a fixed displacement amplitude $A$. 
In the co-moving frame, this corresponds to a modulation of the potential bias within a dimer $\Delta(\tau) = K_0\,\hbar\omega \cos(\omega \tau)$, where $K_0=m\,A\,\omega\,d/\hbar$  is the normalized drive amplitude, with $m$ the mass of the particles and $d$ the distance between the two sites of the dimer \cite{si}.
The atomic state is given by $\ket{\Psi}=\prod_i \ket{\psi_i}$, where $1\leq i \leq N_{\mathrm D}$ with $N_{\mathrm D}$ the number of doubly occupied dimers and \ket{\psi_i} the atomic state on dimer $i$.
We characterize \ket{\Psi} by measuring either the ensemble average of the singlet fraction $\ps=1/N_{\mathrm{tot}} \sum_i|\langle \mathrm s\ket{\psi_i}|^2$ , the triplet fraction $\pt=1/N_{\mathrm{tot}} \sum_i|\langle \mathrm t\ket{\psi_i}|^2$ or the double occupancy $\pdo=1/N_{\mathrm{tot}} \sum_i\left(|\langle \mathrm{D}_+\ket{\psi_i}|^2+|\langle \mathrm{D}_-\ket{\psi_i}|^2\right)$ \cite{Jordens2008, Greif2013a}. The maximal possible values of \pdomax and \psmax are therefore limited by the initial preparation of the system and given as horizontal grey lines in Figs. \ref{fig1} and \ref{fig3b} .

We begin the experiment by characterizing the change in the Floquet state originating from the undriven ground state for a drive frequency $\omega/2\pi= 8$ kHz, larger than both the tunneling amplitude $t/h=548(18)$ Hz and the strength of the on-site interaction $|U|/h$.
This specific frequency is selected to avoid resonant coupling to higher bands of the optical lattice (the first excited band is $h\times 26(3)$ kHz higher in energy) \cite{Weinberg2015}. 
The on-site interaction $U/h$ is set between $- 2.4(2)$ kHz and $2.8(1)$ kHz.
We ramp up the periodic drive in $5$ ms, let the system evolve for $5$ ms, and finally freeze the evolution of the state by raising the potential barrier between the two wells in $100$~$\mu$s.
We measure either \pdo, or \ps and \pt, both with and without the periodic drive.
For the static dimers, \pdo decreases whilst \ps increases when the on-site interactions are varied from attractive to repulsive, as shown in Fig. \ref{fig1}(c).
In the case of the driven dimers, the same qualitative behavior is observed, however, the change in \pdo and \ps with $U$ is much steeper, which can be understood as a consequence of a reduced tunneling.
Furthermore, the periodic drive leads to an increase in the triplet fraction \pt of $0.06(1)$ at most, indicating that most of the atoms remain in the Floquet state connected to the undriven ground state of the double wells.

From a high frequency expansion of the Floquet Hamiltonian, the leading correction to the tunneling in the dimer is given by $t_{\mathrm{eff}} = t\cdot \mathcal{J}_0(K_0)$ where $\mathcal{J}_0$ is the zeroth-order ordinary Bessel function  \cite{Dunlap1986, Lignier2007}.
In Appendix \ref{ap:analytic} we derive the higher order corrections to the tunneling, by performing a high frequency expansion of the periodically driven Hamiltonian \cite{Goldman2014a, Bukov2015b}.
These are proportional to $t^3/\omega^2$ and $U\cdot t^2/\omega^2$ and do not change the observables considerably for the range of interactions and the frequency used in this measurement, as demonstrated by the comparison of analytically and numerically determined effective Hamiltonians in Appendix \ref{ap:numerical}.

To complement our measurement, we also characterize the adiabaticity of the driving scheme by ramping up the drive in $5$ ms, waiting for $5$ ms, reverting the ramp and measuring the final state in the static dimers. 
In the rest of the manuscript, we indicate the results of such a measurement with an $(R)$ superscript.
For drive amplitudes as large as $K_0 = 1.8(3)$, the return fractions \pdor and \psr differ from their original static values by $\Delta p_{\mathrm{DO}}=0.03(2)$ and $\Delta p_{\mathrm{s}}=0.14(4)$, which shows that for a ramp time corresponding to roughly $3\,h/t$ we connect the static and driven Hamiltonians nearly adiabatically \cite{si}.

\begin{figure}[t]
    \includegraphics{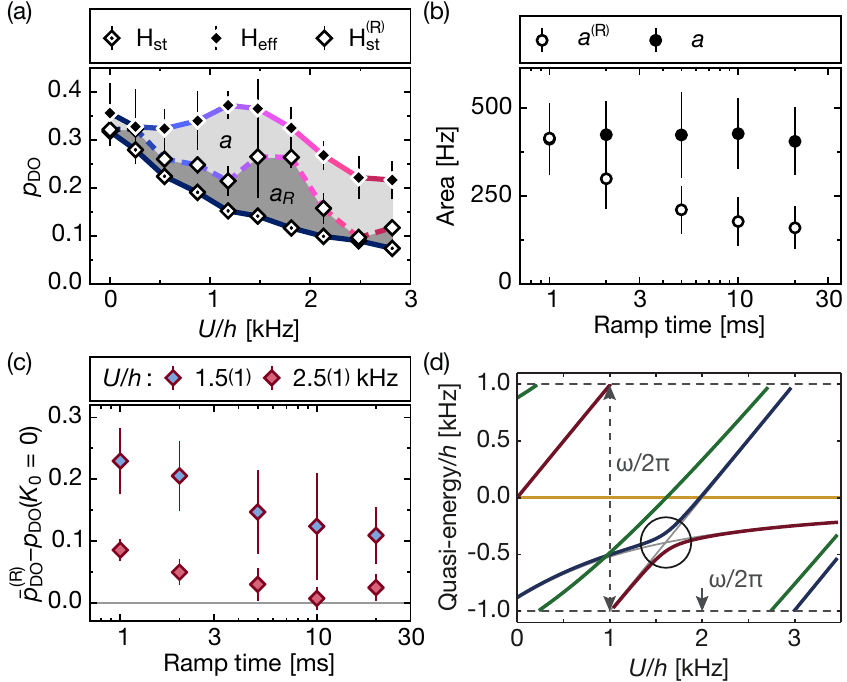}
    \caption{
    Adiabaticity and quasi-energy spectrum of the resonantly driven two-body system ($U\approx\hbar\,\omega$). 
		    (a) An exemplary resonance peak of the double occupancy for a ramp time of 10~ms in the lattice driven with $\omega/2\pi=2$ kHz at $K_0= 1.14(2)$. The data points shown are measured in the driven system \pbdo (filled black diamonds), after reverting the loading ramp \pbdor (open diamonds), and in the static lattice \pdo (open-dotted diamonds). 
		    The variation of color in the connecting lines indicates the changing content of the static eigenstates in the target Floquet state as the interactions are varied.
		    (b) For different ramp times the adiabaticity can be quantified by comparing $a$ and $a^{(R)}$ defined as the area between the static and the other curves respectively. 
		    (c) Two different behaviors are observed in the return fraction $\pbdor - \pdo(K_0=0)$ depending whether the interaction strength is chosen on the resonance peak ($U/h = 1.5(1)$ kHz, blue filled points) or away from it ($U/h = 2.5(1)$ kHz, red points).
		    Error bars in (a,b,c) denote the standard deviation of at least 4 measurements.
		    (d) The time-periodic Hamiltonian is described by a quasi-energy spectrum. 
		    When the amplitude of the drive is zero it is given by the static spectrum modulo $\hbar\omega$ (grey lines). 
		    Switching on a driving amplitude in the resonant case leads to a mixing of the static energy levels \ketstilde and \ketmtilde and creates an avoided crossing (shown for $K_0=0.2$). 
		    The emerging gap in the quasi-energy spectrum is to lowest order given by $4t\cdot\mathcal{J}_1(K_0)$ (see Appendix \ref{ap:analytic}).
    }
	\label{fig2}
\end{figure}

In the previous measurement the leading effect of the periodic drive renormalizes the tunneling, independent of the interaction strength.
To investigate the interplay between interactions and modulation, we select a driving frequency $\omega/2\pi=2$ kHz which can be comparable to $U$. 
In this regime, the periodic drive has been predicted to generate density-dependent tunneling processes \cite{Mentink2015, Bukov2016, Coulthard2016, Kitamura2016}.
At this lower frequency, the micromotion at the time scale of the periodic drive becomes visible. 
First, we concentrate on the slow dynamics governed by the effective Hamiltonian, while the dependence of the micromotion on the interaction strength will be studied further below.
To this end, we remove the fast dynamics by averaging measurements over one modulation cycle \cite{Bukov2014a, si}, and denote the averaged quantities by $\bar{p}$.

When setting the on-site interactions close to the modulation frequency (i.e. $U\approx\hbar\omega$), the resulting Floquet eigenstates differ significantly from their static counterparts, even for weak driving (see Appendix \ref{ap:analytic}). 
This is a particularly interesting regime to study the time-scales required for creating modulation-induced changes in the state of the system without irreversibly driving it out of equilibrium. 
In Fig. \ref{fig2}(a), we show how the double occupancy depends on interactions, for a fixed ramp-time of the modulation amplitude.
When the repulsive on-site interactions are set to values close to $U=\hbar \omega$, more double occupancies are observed than in the static case, with the maximal change in \pbdo around $U/h \approx 1.5$ kHz.
At this interaction strength, the states \ketstilde and \ketmtilde are separated by roughly $\omega/2\pi = 2$ kHz (see Appendix \ref{ap:analytic}).

In order to distinguish the contribution of the effective Hamiltonian from non-adiabatic processes, we also measure the return fraction \pbdor, and compare it to the double occupancies in the initial state.
Contrarily to the off-resonant driving, the initial level of double occupancies cannot be recovered for all interactions.
A peak remains visible around $U/h \approx 1.5$ kHz, although it is significantly less pronounced than in the driven lattice (see Fig. \ref{fig2}(a)).
For a given ramp time, we characterize the response to the driving by calculating the area $a$ between the initial \pdo and its value in the driven lattice, and the area $a^{(R)}$ between the initial \pdo and the return values \pbdor.
While the area $a$ does not depend on the ramp time, the area $a^{(R)}$ decreases for longer ramp times (see Fig. \ref{fig2}(b)).
When setting the interactions away from the resonance, a nearly adiabatic transfer becomes possible for the longer ramp times (see Fig. \ref{fig2}(c)).
This measurement therefore determines how long the ramp time is required to be (for a given distance from the resonance) in order to allow for adiabatic transfer to the Floquet state.
Using a ramp time of $10$ ms, we find values for the interactions left $U/h=0.87(5)$ kHz and right $U/h=2.8(1)$ kHz of the resonance that show a large change in the Floquet states while their return fraction does not differ from its initial value by more than $0.06$. 

\begin{figure}
    \includegraphics{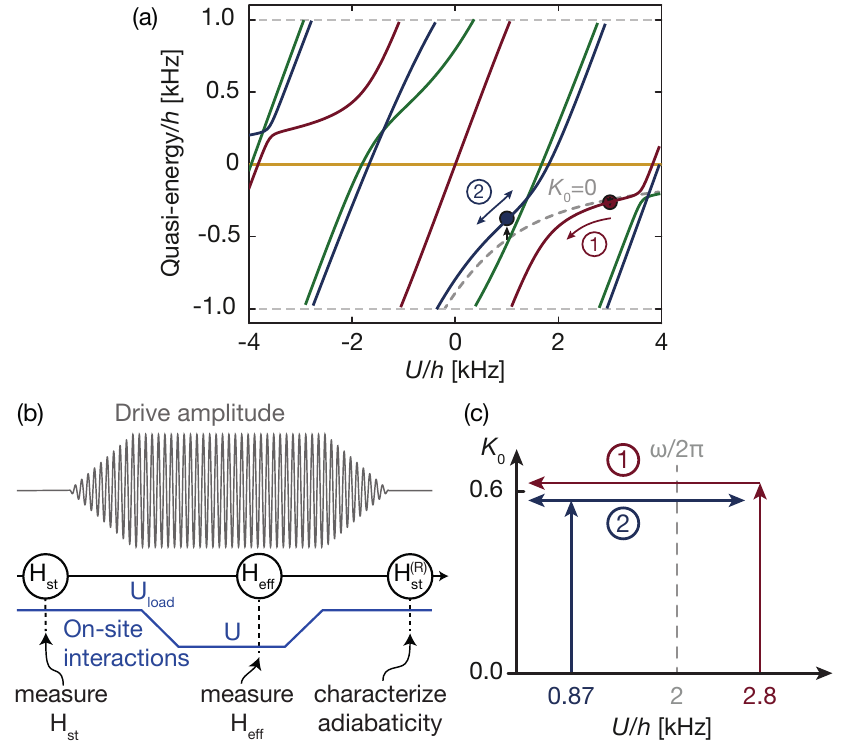}
    \caption{
    Preparation scheme of Floquet states for $U\approx\hbar\,\omega$. 
    (a) Quasi-energy spectrum for $t/h = 445$ Hz, $K_0 =0.6$. Each Floquet state is marked by a different color. 
		The grey dashed line marks the ground state in the absence of modulation. Depending on the initial interaction \Uload when switching on the modulation we connect to a specific Floquet state (indicated with the arrows and circles). We can prepare our system in a single  Floquet state and characterize its nature by measuring the double occupancies and singlets.	
		(b) To enter a specific Floquet state we have to first adiabatically switch on the driving amplitude at a fixed \Uload. In a second step we tune the interactions to $U$ while staying in the effective Hamiltonian. The corresponding trajectories in the $(U, K_0)$ parameter space are schematically illustrated in (c). Once the desired value of $K_0$ is reached, the interaction strength can be freely tuned. Thus, a point in this parameter space can be accessed by two different trajectories.}
	\label{fig3a}
\end{figure}

\begin{figure}
    \includegraphics{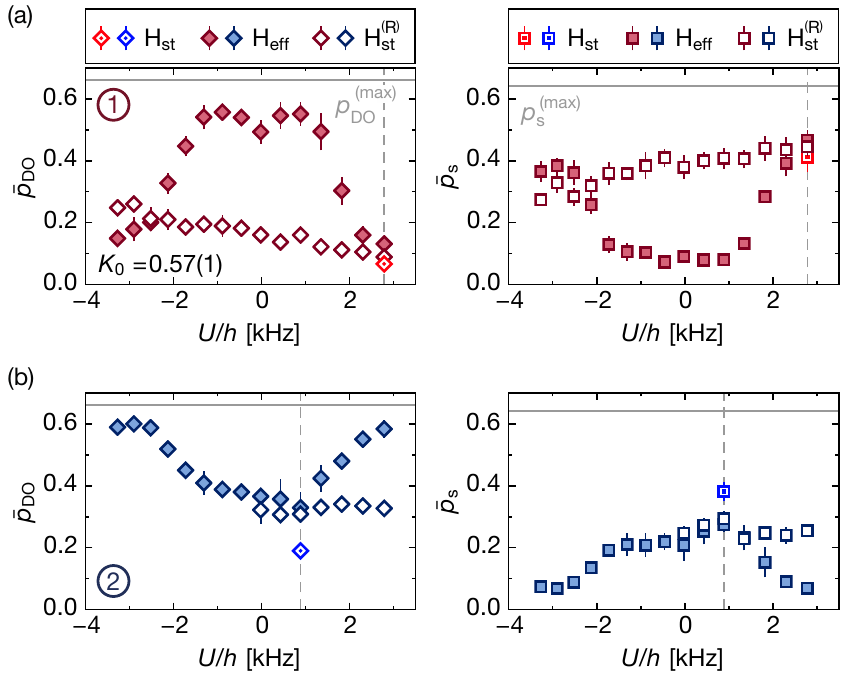}
    \caption{
    Observation of Floquet states for $U\approx\hbar\,\omega$. 
    After preparing at $\Uload=2.8(1)$ kHz (a) and $\Uload=0.87(5)$ kHz (b) (corresponding respectively to trajectories (1) and (2) in Fig. \ref{fig3a} panel (c)), we measure the double occupancy fraction \pbdo and singlet fraction \pbs for the red (a) and blue (b) Floquet states when modulating at 2~kHz with $K_0 = 0.57(1)$.
    Depending on the interaction, the effective states are characterized either by a high double occupancy or singlet fraction depending on the nature of the state. 
    The hollow points indicate the return fraction, obtained by reverting the preparation scheme. 
    The open-dotted data points show the value of the static system at \Uload. 
    Error bars denote the standard deviation of 4 measurements.}
	\label{fig3b}
\end{figure}

The remaining peak in \pbdor can be explained by considering the change in the effective Hamiltonian.
In a driven system, the energy is not conserved, and must be replaced by the quasi-energy, which is only defined modulo $\hbar\omega$.
Thus, two eigenstates of the static Hamiltonian can possess the same quasi-energy, such as the pairs \ketstilde and \ketptilde, or \ketstilde and \ketmtilde, as shown in Fig. \ref{fig2}(d).
When the drive is switched on the first pair is unaffected, as states \ketstilde and \ketptilde are not coupled to each other by the periodic drive.
However, the degeneracy between states \ketstilde and \ketmtilde is lifted as soon the driving amplitude becomes non-zero.
The gap between the two resulting states to leading order in $1/\omega$ is given by $4t\,\mathcal{J}_1(K_0)$, where $\mathcal{J}_1$ is the first order Bessel-function (see Appendix \ref{ap:analytic}).
The drive is thus never perturbative at the level crossing: an avoided crossing forms in the quasi-energy levels, and the system cannot remain in an eigenstate. 
Conversely, by setting $U$ away from the resonance condition, the static and driven states can be connected adiabatically provided that the drive is ramped up sufficiently slowly.
These two regimes are indeed observed experimentally, and shown in Fig. \ref{fig2}(c).

In general, simply ramping up the modulation may not be the fastest protocol for reaching a desired final state with maximal fidelity.
Given the appearance of an avoided crossing, it is preferable to start driving the system off-resonantly	. 
Here, we either start above ($\Uload/h=2.8(1)$ kHz) or below ($\Uload/h=0.87(5)$ kHz) the resonance and ramp in $10$~ms into the corresponding driven state, as illustrated in Fig. \ref{fig3a}(b, c).
Then, in a second step, the interactions are linearly ramped to the desired value of $U$ in $10$ ms and the state can be transferred adiabatically along the avoided crossing (see Fig. \ref{fig3a}(a)), before \pbdo and \pbs are measured (see Fig. \ref{fig3a}(b, c)).
For a given final interaction strength $U$, two distinct Floquet states can be accessed depending on the choice of the initial on-site interaction \Uload.
For example, when $\Uload/h=2.8(1)$ kHz the atomic state is initially \ketstilde. 
As the interactions are decreased while staying in the driven system, the state is first transferred to a doubly-occupied state, and back to \ketstilde when $U<-\hbar\omega$ (see the curves with filled data points in Fig. \ref{fig3b}(a) and Appendix \ref{ap:numerical}).
Correspondingly, \pbs decreases at first with $U$, but is restored to its initial level as $U/h \approx -3$ kHz (see Fig. \ref{fig3b}(b)).
In the other Floquet state obtained with $\Uload/h=0.87(5)$ kHz, the opposite behavior is observed: high double-occupancies for large $|U|$ are connected through a state with a high singlet fraction \pbs.
In the Appendix \ref{ap:numerical} we show a comparison of the experimental results of \pbdo and \pbs with the numerical calculations of these observables in the Floquet states. 
We have therefore observed that despite the final parameters being identical, the state of the system is determined by the path taken to reach these parameters. 
Our procedure of ramping the interaction strength can be generalized by ramping the frequency of the drive, which would therefore be an equivalent route in other physical systems. 

To quantify the fidelity of this preparation protocol, we measure the return fractions \pbdor and \pbsr for each initial interaction strength, by reverting first the interaction ramp, and then the drive ramp.
The change in the return fraction increases smoothly as the interactions are varied, and differ by $0.2$ at most from the initial corresponding quantity.
This indicates that the observed increase of population of the unwanted states is gradual, rather than linked to a closing gap in the effective Hamiltonian. 
In particular, the peak associated to the resonance observed in Fig. \ref{fig2} has vanished using this protocol.
An extension of this scheme could be used to prepare specific excited states of the static double wells by removing the periodic drive at a specific $U$ after crossing the resonance.
For example, by starting at $\Uload/h = 2.8(1)$ kHz, then ramping the interactions to $U/h \approx -3$ kHz, and subsequently ramping down the periodic drive (see Fig. \ref{fig3b}(a), right), a singlet \ketstilde can be prepared at attractive interactions.

\begin{figure}[t]
    \includegraphics{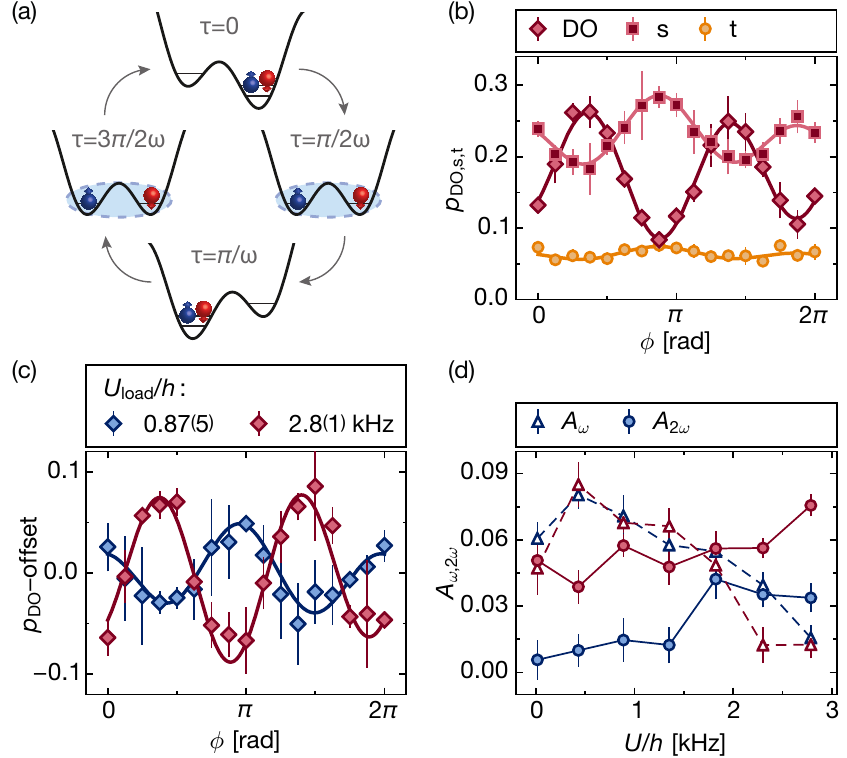}
    \caption{Observation of micromotion in the effective Floquet Hamiltonian. 
(a) During a modulation cycle, the imbalance between the wells reaches a maximum at $\omega \tau_M + \phi =0, \pi$, while it cancels at $\omega \tau_M+\phi = \pi/2, 3\pi/2$. The double occupancy is maximal in the former, while the singlet is favored in the latter, which both occur twice per period. Thus, the micromotion in our observables has a frequency of $2\omega$. 
(b) This is confirmed experimentally by stopping the evolution of the atomic state at different phases $\Phi$ within a full Floquet period and subsequently measuring \ps, \pt and \pdo to observe the micromotion at $U/h = 2.8(1)$ kHz and $K_0 = 1.19(7)$.
(c) The different Floquet states accessed as in Fig. \ref{fig3b} can be distinguished by the phase of their micromotion measured with $K_0 = 1.19(7)$ and final $U/h = 2.8(1)$ kHz. 
(d) Measured micromotion amplitude at frequency $2\omega$ and $\omega$ for the two Floquet states vs interaction $U$ at modulation strength $K_0 = 1.19(7)$. 
Error bars in (b,c) denote the standard deviation of 3 measurements, error bars in (d) are the standard deviation of the amplitude given by a bootstrap method.      
    }
	\label{fig4}
\end{figure}

So far we have averaged our observables over one driving cycle. 
We now turn to analyzing their fast dynamics on sub-cycle timescales which is not captured by the picture of an effective Hamiltonian. 
In our measurement procedure, the tunneling is quenched below $h\times 3$ Hz in $100$ $\mu$s, faster than the oscillation period $2\pi/\omega=500$ $\mu$s, allowing us to probe the fast dynamics of the system.
To this end, we sample different points of the modulation cycle by varying the phase of the periodic drive $\phi$, and always performing the measurement at the same absolute time $\tau_M$.
When the drive is ramped up sufficiently slowly, the launching phase does not play any role in the subsequent evolution \cite{Novicenko2017}, which we have experimentally verified \cite{si}.
In Fig. \ref{fig4}(b), we show  the evolution of the instantaneous \pdo, \ps and \pt, when varying $\phi$ between $0$ and $2\pi$, for $U/h=2.8(1)$ kHz and shaking at frequency  $\omega/2\pi=2$ kHz.
All observables oscillate at twice the driving frequency, with the oscillations in \pdo and \ps being opposite in phase.
This is expected, as during the full drive cycle illustrated in Fig. \ref{fig4}(a), the energy imbalance between the wells is maximal twice ($\omega\tau_M+\phi =0,\,\pi$), and cancels twice ($\omega\tau_M+\phi =\pi/2,\,3\pi/2$), corresponding respectively to the maxima in \pdo and \ps.
Although \ket{\mathrm t} should be unaffected by the periodic drive, a much weaker oscillation is observed in \pt, which can be caused by a residual magnetic field gradient.
The existence of micromotion shows that the effective static Hamiltonian is not sufficient to fully describe the system any more, as the observed time-dependence cannot be ignored for this lower frequency.

We now measure the micromotion for the two Floquet states accessed previously using $\Uload/h=2.8(1)$ kHz or $\Uload/h=0.87(5)$ kHz (see Fig. \ref{fig3b}).
These states are not only differentiated by the averages \pbs and \pbdo, but also by the relative phase of their micromotion, as shown in Fig. \ref{fig4}(c).
We measure the micromotion for both Floquet states as a function of the interaction strength and observe a signal for all $U$(see Fig. \ref{fig4}(d)). 
While the amplitude of the micromotion differs between the two states the overall signal does not depend much on $U$. 
However, as the interaction strength is reduced, an oscillation appears at the driving frequency $\omega$, and even becomes dominant for $U=0$.
This frequency component can be explained by a remaining finite site-offset $\Delta$ or a residual amplitude modulation of the lattice depth (see \cite{si} for more details). 

In conclusion, we have demonstrated the full control of a periodically driven few-level system. 
We could adiabatically connect the initial state to a targeted Floquet state when setting the drive frequency away from any resonant coupling. 
As the drive frequency approached a resonant transition between energy levels, the ramp time required for adiabatic transfer increased, and even diverged directly on the resonance. 
However, the ground state of the static Hamiltonian could nevertheless be adiabatically connected to the desired Floquet state by changing the interaction strength in the driven system.
Furthermore, the local observables developed for static systems could be used directly, by freezing the evolution of the driven state before measuring it. 
Finally, the micromotion on the time scale of the periodic drive was directly visible, confirming the need for a characterization beyond an effective static Hamiltonian.

The versatility of cold atoms experiments offers the possibility to perform similar measurements in a fully connected lattice, and realize a many-body driven system in a future experiment. 
It will therefore be possible to study its dynamics in a controlled setting \cite{Jotzu2015}, and to investigate the possible existence of long-lived quasi-steady states \cite{DAlessio2014a, Lazarides2014, Mori2015, Abanin2015, Kuwahara2016}.

The addition of interactions to periodically driven systems also allows to create novel exotic phases of matter. 
For example, Floquet engineering has been used to create topological states \cite{Jotzu2014,  Aidelsburger2014}, which may lead to new phases of matter in the presence of strong interactions \cite{Zheng2015, Hickey2016, Vanhala2016, Mikami2016}. 
Additionally, the near-resonant modulation shown in this work has been theoretically demonstrated to generate density-dependent hopping, which significantly alters the properties of many-body phases \cite{Mentink2015, Kitamura2016, Coulthard2016}, and could be applied to enhance anti-ferromagnetic interactions in the Hubbard model, or even probe regimes of magnetic order not accessible within this model \cite{Mentink2015, Bermudez2015}. 
Determining the relevant time-scales of the dynamical processes will contribute to the understanding of the scope and limitations of ultrafast optical manipulation of magnetic order \cite{Kirilyuk2010}. 
Finally, the observation of micromotion also paves the way to realizing novel states of matter, which are exclusive to periodically driven systems such as exotic topological states \cite{Rudner2013, Potter2016, VonKeyserlingk2016a} and time crystals \cite{Zhang2017, Choi2017}.

\begin{acknowledgments}
We thank D. Abanin, L. Corman, J. Coulthard, E. Demler, D. Jaksch, A. Lazarides, Y. Murakami, N. Tsuji and  P. Werner for insightful discussions, and are most grateful to W. Zwerger for his insights and a careful reading of the manuscript. T. Esslinger acknowledges hospitality by Keble College, Oxford and the University of Oxford. We acknowledge SNF, NCCR-QSIT, QUIC (Swiss State Secretary for Education, Research and Innovation contract number 15.0019) and SQMS (ERC advanced grant) for funding. 
\end{acknowledgments}

\renewcommand{\theequation}{A\arabic{equation}}
\makeatletter
\renewcommand{\thefigure}{A\@arabic\c@figure} 
\renewcommand{\thetable}{A\@arabic\c@table} 
\setcounter{figure}{0}  

\appendix
\section{Analytic treatment of the periodically modulated double well}
\label{ap:analytic}

In the following we provide an analytic description of the periodically modulated double well system. First, the static Hamiltonian and its properties are discussed. Afterwards, the driven system is treated with a high frequency expansion approach. In this context, we derive explicit expressions for the effective static Hamiltonian and the time dynamics of the system within one period for both an off-resonant and a near resonant modulation of the double wells.

\subsection{Static double well}
We consider two distinguishable Fermions with spin $\uparrow$ and $\downarrow$ on a double well. The starting point is a continuum Hamiltonian for Fermions with two spin states $\sigma$ and zero range interactions $V_{\mathrm{int}}(\mathbf{r})=4\pi a/m \delta(\mathbf{r})$, where $m$ is the mass of the atoms and $a$ the s-wave scattering length (we set $\hbar=1$). The tight-binding Hubbard Hamiltonian is obtained upon replacing the field operators with $\hat{\Psi}^{\dagger}_{\sigma}(\mathbf{r})=\sum_{l=\mathrm{L,R}}w_l(\mathbf{r})c^{\dagger}_{l\sigma}$. Here, $c^{\dagger}_{\mathrm{L}\sigma}$ ($c^{\dagger}_{\mathrm{R}\sigma}$) denote the fermionic creation operators for a particle with spin $\sigma$ on the left (right) side and $w_{\mathrm{L}}(\mathbf{r})$ (respectively $w_{\mathrm{R}}(\mathbf{r})$) are the (real) Wannier functions of the underlying extended lattice, which are determined as eigenstates of the band projected position operator \cite{Uehlinger2013}.

We choose to work in the Fock basis
\begin{equation}
\begin{aligned}
&\left|\uparrow\downarrow,0\right\rangle &=\ \ & c^{\dagger}_{\mathrm{L}\downarrow}c^{\dagger}_{\mathrm{L}\uparrow}\left|0\right\rangle \\
&\left|\uparrow,\downarrow\right\rangle &=\ \ & c^{\dagger}_{\mathrm{R}\downarrow}c^{\dagger}_{\mathrm{L}\uparrow}\left|0\right\rangle \\
&\left|\downarrow,\uparrow\right\rangle &=\ \ & c^{\dagger}_{\mathrm{R}\uparrow}c^{\dagger}_{\mathrm{L}\downarrow}\left|0\right\rangle \\
&\left|0,\uparrow\downarrow\right\rangle &=\ \ & c^{\dagger}_{\mathrm{R}\downarrow}c^{\dagger}_{\mathrm{R}\uparrow}\left|0\right\rangle
\end{aligned}
\label{FockBasis}
\end{equation}
in which the Hamiltonian takes the form
\begin{equation}
H_0=\begin{pmatrix} U & -t-\delta t & t+\delta t & V_{\mathrm{ct}}\\
                   -t-\delta t & V_{\mathrm{nn}} & -V_{\mathrm{de}} & -t-\delta t\\
								   t+\delta t & -V_{\mathrm{de}} & V_{\mathrm{nn}} & t+\delta t\\
								   V_{\mathrm{ct}} & -t-\delta t & t+\delta t & U \end{pmatrix}
\end{equation}
Here, the tunneling amplitude $t$ and the on site interaction $U$ are given by
\begin{eqnarray}
t&=&-\int d^3r\,w_{\mathrm{L}}(\mathbf{r})\left[-\frac{1}{2m}\nabla^2+V(\mathbf{r})\right]w_{\mathrm{R}}(\mathbf{r})\\
U &=& \frac{4\pi a}{m}\int d^3r\,\left|w_{\mathrm{L,R}}(\mathbf{r})\right|^4
\end{eqnarray}
with the lattice potential $V(\mathbf{r})$ \cite{si}. Since for this work, there is no static site offset between the two sites, the Wannier functions are symmetric around the center of the wells $w_{\mathrm{L}}(\mathbf{r})=w_{\mathrm{R}}(\mathbf{-r})$ and the interactions are equal on both sides.

The other terms appearing in the Hamiltonian are higher band corrections, namely the correlated tunneling $V_{\mathrm{ct}}$ describing the hopping of atom pairs, the nearest-neighbor interaction $V_{\mathrm{nn}}$ and the direct spin exchange $V_{\mathrm{de}}$ connected to spin flips between the two Fermions on adjacent sites. In the two-site problem all of these terms are equal and obtained as
\begin{equation}
V_{\mathrm{ct}} = V_{\mathrm{nn}}=V_{\mathrm{de}}=\frac{4\pi a}{m}\int d^3r\,\left|w_{\mathrm{L}}(\mathbf{r})\right|^2\left|w_{\mathrm{R}}(\mathbf{r})\right|^2
\label{V}
\end{equation}
Furthermore, there is a density assisted hopping term $\delta t$, which accounts for a correction to the tunneling amplitude when there is another particle of opposite spin present in the double well. It is given by
\begin{equation}
\delta t = -\frac{4\pi a}{m}\int d^3r\,\left|w_{\mathrm{L,R}}(\mathbf{r})\right|^2w_{\mathrm{L}}(\mathbf{r})w_{\mathrm{R}}(\mathbf{r})
\label{Deltat}
\end{equation}
All of these latter corrections are small for the static lattices used in this paper. In order to estimate their magnitude, we use the same method as for the calculation of $K_0$ and consider a cut through the lattice potential at $y=z=0$ \cite{si}. We then make the approximation that the lattice potential is separable in the x- and z-directions, determine the Wannier functions in this one-dimensional problem as eigenstates of the band-projected position operator \cite{Uehlinger2013} and calculate the corrections according to Eqs.\:(\ref{V}) and (\ref{Deltat}). We find that $V_{\mathrm{ct}}/U,V_{\mathrm{nn}}/U,V_{\mathrm{de}}/U\approx 10^{-3}$ and $\delta t/U\approx 10^{-2}$.

Therefore, we can first restrict the discussion to the case where we only have a tunneling $t$ and an on site interaction $U$ and the Hamiltonian is given by
\begin{equation}
H_0=\begin{pmatrix} U & -t & t & 0\\
                   -t & 0 & 0 & -t\\
								   t & 0 & 0 & t\\
								   0 & -t & t & U \end{pmatrix}
\label{SimpleStatic}
\end{equation}
In this case, it is convenient to change to a new basis which consists of a singlet state $\left|\mathrm{s}\right\rangle$, a triplet state $\left|\mathrm{t}\right\rangle$ and two states containing double occupancies $\left|\mathrm{D}_{\pm}\right\rangle$ given by
\begin{equation}
\begin{aligned}
&\left|\mathrm{t}\right\rangle &=\ \ & \frac{1}{\sqrt{2}}\left(\left|\uparrow,\downarrow\right\rangle + \left|\downarrow,\uparrow\right\rangle\right) \\
&\left|\mathrm{D}_+\right\rangle &=\ \ & \frac{1}{\sqrt{2}}\left(\left|\uparrow\downarrow,0\right\rangle + \left|0,\uparrow\downarrow\right\rangle\right)\\
&\left|\mathrm{D}_-\right\rangle &=\ \ & \frac{1}{\sqrt{2}}\left(\left|\uparrow\downarrow,0\right\rangle - \left|0,\uparrow\downarrow\right\rangle\right) \\
&\left|\mathrm{s}\right\rangle &=\ \ & \frac{1}{\sqrt{2}}\left(\left|\uparrow,\downarrow\right\rangle - \left|\downarrow,\uparrow\right\rangle\right)
\end{aligned}
\label{SingletBasis}
\end{equation}
In this new basis, the Hamiltonian takes the simple form
\begin{equation}
H^{'}_0=\begin{pmatrix} 0 & 0 & 0 & 0 \\
                  0 & U & 0 & -2t \\
									0  & 0 & U & 0 \\
									0 & -2t & 0 & 0 \end{pmatrix}
\label{SimpleStaticBasis}
\end{equation}
We see that $\left|\mathrm{t}\right\rangle$ and $\left|\mathrm{D}_{-}\right\rangle$ are eigenstates with energies $0$ and $U$, respectively. The other two eigenstates are superpositions of $\left|\mathrm{s}\right\rangle$ and $\left|\mathrm{D}_{+}\right\rangle$ with eigenenergies
\begin{equation}
E_{1,4}=\frac{1}{2}\left(U\mp\sqrt{16 t^2+U^2}\right)
\end{equation}
The spectrum is shown in Fig. \ref{fig1} in the main text. For large repulsion, the singlet is the ground state, whereas for strong attractive interactions it is the $\left|\mathrm{D}_{+}\right\rangle$ state containing double occupancies.

\subsection{Periodically modulated system}
Now we add the periodic modulation $V(\tau)$ to the Hamiltonian $H_0$, such that the total Hamiltonian in the lab frame has the form
\begin{equation}
H_{\mathrm{lab}}(\tau)=H_0+V(\tau)
\label{Hlab}
\end{equation}
The time-dependent part $V(\tau)$ is given by
\begin{equation}
V(\tau)=\Delta(\tau) h_{\Delta}
\label{DriveOp}
\end{equation}
where $\Delta(\tau)=\omega K_0 \cos(\omega\tau)$ is the modulated site offset expressed by the dimensionless shaking amplitude $K_0$ and the driving frequency $\omega$. The operator for the site offset $h_{\Delta}$ is most intuitively expressed in the Fock basis (\ref{FockBasis})
\begin{equation}
h_{\Delta}=\begin{pmatrix} 1 & 0 & 0 & 0 \\
                  0 & 0 & 0 & 0 \\
									0  & 0 & 0 & 0 \\
									0 & 0 & 0 & -1 \end{pmatrix}
\end{equation}

We will treat this Hamiltonian with the Floquet approach, where the time evolution operator from an initial time $\tau_i$ to a final time $\tau_f$ can be written as \cite{Goldman2014a, Bukov2015b}
\begin{equation}
U(\tau_f,\tau_i)=e^{-iK(\tau_f)}e^{-i(\tau_f-\tau_i)H_{\mathrm{eff}}}e^{iK(\tau_i)}
\label{EvolOp}
\end{equation}
Here, the effective Hamiltonian $H_{\mathrm{eff}}$ is time independent and governs the long term dynamics of the system, while the time periodic kick operator $K(\tau)=K(\tau+T)$ describes the evolution within one period of the drive (the so called `micromotion'). Note that we choose to work with the non-stroboscopic approach where the effective Hamiltonian and kick operator do not depend on the starting phase of the modulation.

The effective Hamiltonian and kick operator can be calculated perturbatively in a high frequency expansion according to \cite{Goldman2014a, Bukov2015b}
\begin{equation}
H_{\mathrm{eff}}=\sum_{n=0}^\infty H_{\mathrm{eff}}^{(n)},\ K(\tau)=\sum_{n=1}^\infty K^{(n)}(\tau)
\label{HFE}
\end{equation}
where the operators are expanded in powers of the inverse frequency $H_{\mathrm{eff}}^{(n)}\propto \omega^{-n}$ and $K^{(n)}(\tau)\propto \omega^{-n}$.

\subsubsection{Rotating frame}
In the following we will consider the strong driving regime where the driving strength $K_0$ is of order $1$. Since in this case the amplitude of the modulation in (\ref{DriveOp}) becomes large in the high frequency limit, we go to a rotating frame via the unitary transformation
\begin{equation}
R_1(\tau)=\exp\left[-i\int{V(\tau)\mathrm{d}\tau}\right]=\exp\left[-i K_0 \sin(\omega \tau) h_{\Delta}\right]
\label{R1}
\end{equation}
The Hamiltonian is transformed according to
\begin{eqnarray}
H_{\mathrm{rot}}(\tau)&=&R_1^{\dagger}(\tau)H_{\mathrm{lab}}(\tau)R_1(\tau)-i R_1^{\dagger}(\tau)\frac{\partial}{\partial\tau}R_1(\tau)\notag \\
				&=&R_1^{\dagger}(\tau)H_0 R_1(\tau)
\label{Hrot}
\end{eqnarray}
while all observables $\hat{O}$ and states $\left|\psi(\tau)\right\rangle$ in the rotating frame are given by
\begin{eqnarray}
\hat{O}_{\mathrm{rot}}(\tau) &=& R_1^{\dagger}(\tau)\hat{O}_{\mathrm{lab}} R_1(\tau) \\
\left|\psi_{\mathrm{rot}}(\tau)\right\rangle &=& R_1^{\dagger}(\tau)\left|\psi_{\mathrm{lab}}(\tau)\right\rangle
\end{eqnarray}

The time evolution operator (\ref{EvolOp}) in the rotating frame can be written as
\begin{equation}
U_{\mathrm{rot}}(\tau_f,\tau_i)=e^{-iK_{\mathrm{rot}}(\tau_f)}e^{-i(\tau_f-\tau_i)H_{\mathrm{eff}}}e^{iK_{\mathrm{rot}}(\tau_i)}
\end{equation}
Note that the exact effective Hamiltonian is the same in the lab and the rotating frame, while the relation between the kick operators is given by
\begin{equation}
e^{-iK(\tau)}=R_1(\tau)e^{-iK_{\mathrm{rot}}(\tau)}
\label{MicroOp}
\end{equation}

For the time independent effective Hamiltonian $H_{\mathrm{eff}}$ we can then find eigenstates $\left|v\right\rangle$ and eigenvalues $\epsilon_v$ with 
\begin{equation}
H_{\mathrm{eff}}\left|v\right\rangle=\epsilon_v \left|v\right\rangle\;\text{and}\;\left|v(\tau)\right\rangle=\exp(-i\epsilon_v \tau)\left|v\right\rangle
\label{Defv}
\end{equation}
The eigenvalues $\epsilon_v$ are called quasi-energies in analogy to Bloch's theorem and are only defined up to multiples of $\omega$, such that they can be restricted to the first Floquet zone given by $-\omega/2<\epsilon<\omega/2$.
In order to construct the eigenstates of the original time periodic Hamiltonian in the lab frame (\ref{Hlab}), we apply the micromotion operator (\ref{MicroOp})
\begin{equation}
\left|v_{\mathrm{lab}}(\tau)\right\rangle=e^{-iK(\tau)}\left|v(\tau)\right\rangle=e^{-i\epsilon_v \tau}R_1(\tau)e^{-iK_{\mathrm{rot}}(\tau)}\left|v\right\rangle
\label{vLab}
\end{equation}
which can also be written as
\begin{equation}
\left|v_{\mathrm{lab}}(\tau)\right\rangle=e^{-i\epsilon_v \tau}\left|\phi(\tau)\right\rangle
\end{equation}
with a time periodic state $\left|\phi(\tau)\right\rangle=\left|\phi(\tau+T)\right\rangle$, which is analogous to the form of Bloch waves in a spatially periodic system.

The time dependent expectation value of an observable $\hat{O}$ measured in such an instantaneous eigenstate is given by
\begin{equation}
\langle \hat{O}\rangle_v(\tau)=\left\langle v\right|e^{iK_{\mathrm{rot}}(\tau)}R^{\dagger}_1(\tau) \hat{O}_{\mathrm{lab}} R_1(\tau)e^{-iK_{\mathrm{rot}}(\tau)}\left|v\right\rangle
\label{TimeDepObs}
\end{equation}
Therefore, the expectation value of an observable may oscillate at the same frequency as the drive. In general, there are two contributions to this micromotion. The first one originates from the kick operator in the rotating frame $K_{\mathrm{rot}}(\tau)$ whose amplitude scales as $1/\omega$ by construction (\ref{HFE}). Hence, this contribution vanishes at high frequency. Second, the transformation from the lab frame to the rotating frame $R_1(\tau)$ (\ref{R1}) also induces an oscillation which is present as long as the driving amplitude is non-zero, even for infinite frequency. 

All observables measured in the main section of the paper (singlet and triplet fractions, double occupancy) are the same in both the lab frame and the rotating frame, \emph{i.e.} $\hat{O}_{\mathrm{rot}}(\tau) = \hat{O}_{\mathrm{lab}}$. Hence, the micromotion shown in Fig. \ref{fig4} only originates from the kick operator $K_{\mathrm{rot}}(\tau)$, and would not be visible for larger driving frequencies.

Note that the time average of the observable over one oscillation period $\overline{\langle \hat{O}\rangle(\tau)}=1/T\int_0^{\tau} \mathrm{d}\tau\:\langle \hat{O}\rangle(\tau)$ is not necessarily equal to the expectation value of the observable in an eigenstate of the time independent effective Hamiltonian $\left\langle v\right|\hat{O}_{\mathrm{lab}}\left|v\right\rangle$, since in general
\begin{equation}
\overline{e^{iK_{\mathrm{rot}}(\tau)}R^{\dagger}_1(\tau) \hat{O}_{\mathrm{lab}} R_1(\tau)e^{-iK_{\mathrm{rot}}(\tau)}}\neq \hat{O}_{\mathrm{lab}}
\end{equation}

We derive the effective Hamiltonian and kick operator in the high frequency expansion (\ref{HFE})
\begin{equation}
H_{\mathrm{eff}}=\sum_{n=0}^\infty H_{\mathrm{eff,rot}}^{(n)},\ K_{\mathrm{rot}}(\tau)=\sum_{n=1}^\infty K_{\mathrm{rot}}^{(n)}(\tau)
\label{HFErot}
\end{equation}
In our notation we make explicit that the individual summands of the expansion $H_{\mathrm{eff,rot}}^{(n)}$ are different from the ones in the lab frame (\ref{HFE}), even though the full effective Hamiltonian $H_{\mathrm{eff}}$ is identical in both frames.

In the following, we consider two different cases: First, we discuss the off-resonant modulation, where the driving frequency is much larger than the static parameters of the system ($\omega\gg U,t$). In a second step, we will treat the resonant case where $\omega\approx U\gg t$.

\subsubsection{Off-resonant shaking}
If the driving frequency is much larger than the tunneling and interaction $\omega\gg U,t$ the effective Hamiltonian in the rotating frame (\ref{Hrot}) reads
\begin{equation}
H_{\mathrm{rot}}(\tau)=\begin{pmatrix} U & -t(\tau) & t(\tau) & 0\\
                   -t^*(\tau) & 0 & 0 & -t(\tau)\\
								   t^*(\tau) & 0 & 0 & t(\tau)\\
								   0 & -t^*(\tau) & t^*(\tau) & U \end{pmatrix}
\end{equation}
where $(...)^*$ denotes the complex conjugation. In the Hamiltonian, the time dependent site offset in the lab frame has been converted to a time dependent phase of the tunnelings
\begin{equation}
t(\tau)=t\exp\left[i K_0\sin(\omega\tau)\right]
\end{equation}
Performing the expansion (\ref{HFErot}), we find that the effective Hamiltonian is to lowest order given by
\begin{equation}
H_{\mathrm{eff,rot}}^{(0)}=\begin{pmatrix} U & -t \mathcal{J}_0(K_0) & t\mathcal{J}_0(K_0) & 0\\
                   -t\mathcal{J}_0(K_0) & 0 & 0 & -t\mathcal{J}_0(K_0)\\
								   t\mathcal{J}_0(K_0) & 0 & 0 & t\mathcal{J}_0(K_0)\\
								   0 & -t\mathcal{J}_0(K_0) & t\mathcal{J}_0(K_0) & U \end{pmatrix}
\label{LowOrderOffRes}
\end{equation}
which describes the renormalization of the static tunneling $t$ by a 0-th order Bessel function $\mathcal{J}_0(K_0)$ (compare to the static Hamiltonian (\ref{SimpleStatic})). The spectrum and the eigenstates to lowest order can therefore be obtained from the ones of the static Hamiltonian $H_0$ by replacing $t\longrightarrow t\mathcal{J}_0(K_0)$.

The next order proportional to $1/\omega$ vanishes identically $H_{\mathrm{eff}}^{(1)}=\mathbf{0}$ and the leading corrections are obtained from $H_{\mathrm{eff}}^{(2)}$. It contains several new terms which are not present in the static Hamiltonian. They are listed in Table\:\ref{CorrOffRes} up to terms containing Bessel functions $\mathcal{J}_n(K_0)$ with $n\leq 1$. The corrections start to matter for the largest interaction that was used in the experiment as far as the spectrum is concerned, in particular the energy of the lowest energy state crosses the triplet energy for $U/h\approx \pm 2250$ Hz (see discussion in Appendix B and Fig.\:\ref{fig:ThVsNum_sp}). However, the influence on the observables double occupancy and singlet fraction are not very pronounced, see Fig.\:\ref{fig:ThVsNum}.

\begin{table}[htb]
	\begin{center}
		\begin{tabular}{ | l | c | c |}
			\hline
			Quantity & $1/\omega^2$ correction & Value (U/h=3000 Hz) \\ \hline
			$t$ & $-4t^3/\omega^2\mathcal{J}_0(K_0)\mathcal{J}^2_1(K_0)$ & $-h\times 1.2$ Hz \\ \hline
			$U$ & $-4t^2 U/\omega^2\mathcal{J}^2_1(K_0)$ & $-h\times 19.0$ Hz \\ \hline
			$V_{\mathrm{nn}}$,$V_{\mathrm{de}}$,$V_{\mathrm{ct}}$ & $4t^2 U/\omega^2\mathcal{J}^2_1(K_0)$ & $h\times 19.0$ Hz \\ \hline
		\end{tabular}
	\end{center}
	\caption{Summary of the leading corrections to the lowest order expansion of the effective Hamiltonian Eq.\:(\ref{LowOrderOffRes}) in the off-resonant case. Terms containing Bessel functions $\mathcal{J}_n(K_0)$ with $n>1$ were omitted. The last column gives the values of the correction in Hz for the largest interaction $U/h=3000\:\mathrm{Hz}$ that was used in the experiment (for $t/h = 548$ Hz, $K_0=1.8$).}
	\label{CorrOffRes}
\end{table}

The two lowest orders of the kick operator are given by
\begin{equation}
K_{\mathrm{rot}}^{(1)}(\tau) = 2i\frac{t}{\omega}\mathcal{J}_1(K_0)\cos(\omega\tau)\begin{pmatrix} 0 & 1 & -1 & 0\\
                   -1 & 0 & 0 & 1\\
								   1 & 0 & 0 & -1\\
								   0 & -1 & 1 & 0 \end{pmatrix}
\label{Koffres1}
\end{equation}
and
\begin{eqnarray}
&K_{\mathrm{rot}}^{(2)}(\tau) = 2\frac{tU}{\omega^2}\mathcal{J}_1(K_0)\sin(\omega\tau)\begin{pmatrix} 0 & 1 & -1 & 0 \notag\\
                   1 & 0 & 0 & -1\\
								   -1 & 0 & 0 & 1\\
								   0 & -1 & 1 & 0 \end{pmatrix}\\
							&+ 8\frac{t^2}{\omega^2}\mathcal{J}_0(K_0)\mathcal{J}_1(K_0)\sin(\omega\tau)\begin{pmatrix} 1 & 0 & 0 & 0\\
                   0 & 0 & 0 & 0\\
								   0 & 0 & 0 & 0\\
								   0 & 0 & 0 & 1 \end{pmatrix}
\end{eqnarray}
It becomes apparent that to leading order the micromotion amplitude is determined by the ratio $t\mathcal{J}_1(K_0)/\omega$, while for the next terms the ratio $U/\omega$ also becomes important.

\subsubsection{Near-resonant shaking}
Now we turn to the resonant case where $\omega\approx U\gg t$ in the strong driving regime. Here, not only the amplitude of the modulation becomes large in the high frequency limit but also the interaction term proportional to $U$. Therefore, in addition to the transformation (\ref{R1}) we perform a second rotation according to \cite{Goldman2015}
\begin{equation}
R_2(\tau)=\exp\left[-i\omega\tau h_{U}\right]
\label{R2}
\end{equation}
where the interaction operator $h_{U}$ is given by
\begin{equation}
h_{U}=\begin{pmatrix} 1 & 0 & 0 & 0 \\
                  0 & 0 & 0 & 0 \\
									0  & 0 & 0 & 0 \\
									0 & 0 & 0 & 1 \end{pmatrix}
\end{equation}
We then follow the discussion of the off-resonant case, replacing the operator $R_1(\tau)$ by the product $R(\tau)=R_2(\tau)R_1(\tau)$. The Hamiltonian (\ref{Hlab}) in this rotating frame thus becomes
\begin{equation}
H_{\mathrm{rot}}(\tau)=\begin{pmatrix} U-\omega & -t_+(\tau) & t_+(\tau) & 0\\
                   -t_+^*(\tau) & 0 & 0 & -t_-(\tau)\\
								   t_+^*(\tau) & 0 & 0 & t_-(\tau)\\
								   0 & -t_-^*(\tau) & t_-^*(\tau) & U-\omega \end{pmatrix}
\end{equation}
with
\begin{equation}
t_{\pm}(\tau)=t\exp\left[i(\pm\omega\tau+K_0\sin(\omega\tau))\right]
\end{equation}
Again, the oscillating site offset has been converted to a phase factor for the tunneling. In addition, the second transformation adds another phase whose sign depends on which of the two states containing a double occupancy is involved in the tunneling process. Furthermore, the interaction $U$ has been replaced by the detuning from the resonance $\delta=\omega-U$, such that we can perform again a high frequency expansion (\ref{HFErot}) in the two small parameters $t/\omega$ and $\delta/\omega$.

For the resonant case, the effective Hamiltonian to lowest order is given by
\begin{equation}
H_{\mathrm{eff,rot}}^{(0)}=\begin{pmatrix} U-\omega & t \mathcal{J}_1(K_0) & -t\mathcal{J}_1(K_0) & 0\\
                   t\mathcal{J}_1(K_0) & 0 & 0 & -t\mathcal{J}_1(K_0)\\
								   -t\mathcal{J}_1(K_0) & 0 & 0 & t\mathcal{J}_1(K_0)\\
								   0 & -t\mathcal{J}_1(K_0) & t\mathcal{J}_1(K_0) & U-\omega \end{pmatrix}
\label{LowOrderRes}
\end{equation}
Unlike in the off-resonant case, the tunneling matrix elements scale with the first order Bessel function $t\mathcal{J}_1(K_0)$. This can be interpreted as density assisted tunneling introduced by the resonant modulation, since the hopping process can only occur if a particle has a neighbor on the adjacent side with which it interacts.

Also note the important sign change for the tunneling matrix elements compared to the static Hamiltonian (\ref{SimpleStatic}) and the off-resonant modulation (\ref{LowOrderOffRes}). As a consequence, the singlet now couples to the state which is adiabatically connected to the state $\left|\mathrm{D}_-\right\rangle$ in the static Hamiltonian rather than $\left|\mathrm{D}_+\right\rangle$. This becomes evident if we write the Hamiltonian (\ref{LowOrderRes}) in the basis (\ref{SingletBasis}) which yields
\begin{equation}
H_{\mathrm{eff,rot}}^{'(0)}=\begin{pmatrix} 0 & 0 & 0 & 0 \\
                  0 & U-\omega  & 0 & 0 \\
									0  & 0 & U-\omega  & 2t\:\mathcal{J}_1(K_0) \\
									0 & 0 & 2t\:\mathcal{J}_1(K_0) & 0 \end{pmatrix}
\end{equation}
Comparing this to the static Hamiltonian (\ref{SimpleStaticBasis}) shows that the singlet is coupled to the other double occupancy state $\left|\mathrm{D}_-\right\rangle$ around the resonance, which leads to the opening of a gap of size $4t\:\mathcal{J}_1(K_0)$ (see Fig.\:\ref{fig2}(d)).

Finally, notice that the interaction has been replaced by $\delta=\omega-U$. Apart from the convergence criterion of the high frequency expansion which was mentioned before, this also has the physical consequence that the sign of the detuning determines whether the system effectively exhibits an attractive or repulsive interaction.

The higher order corrections to $H_{\mathrm{eff}}$ up to terms $1/\omega^2$ are listed in Table\:\ref{CorrRes} including terms containing Bessel functions $\mathcal{J}_n(K_0)$ with $n\leq 1$. Unlike in the off-resonant case, the first order proportional to $1/\omega$ does not vanish. In fact, it reproduces the Schrieffer-Wolff transformation for the case $K_0=0$ and $\omega=U$ which allows to describe the Hubbard model with an effective spin Heisenberg model in the limit of large interactions $U\gg t$ \cite{Bukov2015b}. The second order proportional to $1/\omega^2$ makes it apparent that the series is an expansion in the two small parameters $t/\omega$ and $\delta/\omega=(\omega-U)/\omega$ and it breaks down if the detuning from the resonance is too large (see Fig. \ref{fig:ThVsNum_sp}).

\begin{table}[htb]
	\begin{center}
		\begin{tabular}{ | l | c | c |}
			\hline
			Quantity & $1/\omega$ \\ \hline
			$t_{\pm}$ & - \\ \hline
			$U-\omega$ & $t^2/\omega [2\mathcal{J}^2_0(K_0)+\mathcal{J}^2_1(K_0)]$ \\ \hline
			$V_{\mathrm{nn}}$,$V_{\mathrm{de}}$ & $-t^2/\omega [2\mathcal{J}^2_0(K_0)+\mathcal{J}^2_1(K_0)]$ \\ \hline
			$V_{\mathrm{ct}}$ & $t^2/\omega [2\mathcal{J}^2_0(K_0)-\mathcal{J}^2_1(K_0)]$ \\ \hline
		\end{tabular}
		\begin{tabular}{ | l | c | c |}
			\hline
			Quantity & $1/\omega^2$ \\ \hline
			$t_{\pm}$ & $\pm t^3/\omega^2\mathcal{J}_1(K_0)) [2\mathcal{J}^2_0(K_0)+\mathcal{J}^2_1(K_0)]$ \\ \hline
			$U-\omega$ & $t^2/(2\omega^2)(\omega-U) [4\mathcal{J}^2_0(K_0)+\mathcal{J}^2_1(K_0)]$ \\ \hline
			$V_{\mathrm{nn}}$,$V_{\mathrm{de}}$ & $-t^2/(2\omega^2)(\omega-U) [4\mathcal{J}^2_0(K_0)+\mathcal{J}^2_1(K_0)]$ \\ \hline
			$V_{\mathrm{ct}}$ & $t^2/(2\omega^2)(\omega-U) [4\mathcal{J}^2_0(K_0)-\mathcal{J}^2_1(K_0)]$ \\ \hline
		\end{tabular}
	\end{center}
	\caption{Summary of the leading corrections to the lowest order expansion of the effective Hamiltonian Eq.\:(\ref{LowOrderRes}) in the resonant case. Corrections containing Bessel functions $\mathcal{J}_n(K_0)$ with $n>1$ were omitted. The terms proportional to $1/\omega$ reproduce the Schrieffer-Wolff transformation for the case $K_0=0$ and $\omega=U$.}
	\label{CorrRes}
\end{table}

Finally, the first order in the expansion of the kick operator is given by
\begin{equation}
\begin{aligned}
&K_{\mathrm{rot}}^{(1)}(\tau) = i\frac{t}{\omega}\mathcal{J}_0(K_0)
								 \begin{pmatrix} 0 & e^{i\omega\tau} & -e^{i\omega\tau} & 0\\
                   -e^{-i\omega\tau} & 0 & 0 & -e^{-i\omega\tau}\\
								   e^{-i\omega\tau} & 0 & 0 & e^{-i\omega\tau}\\
								   0 & e^{i\omega\tau} & -e^{i\omega\tau} & 0 \end{pmatrix} \\
							&+i\frac{t}{2\omega}\mathcal{J}_1(K_0)
								\begin{pmatrix} 0 & e^{2i\omega\tau} & -e^{2i\omega\tau} & 0\\
                   -e^{-2i\omega\tau} & 0 & 0 & e^{-2i\omega\tau}\\
								   e^{-2i\omega\tau} & 0 & 0 & -e^{-2i\omega\tau}\\
								   0 & -e^{2i\omega\tau} & e^{2i\omega\tau} & 0 \end{pmatrix}
\end{aligned}
\end{equation}
As in the off-resonant case (\ref{Koffres1}), the micromotion amplitude is to lowest order determined by the ratio $t\mathcal{J}_1(K_0)/\omega$. 
The first term in the kick operator proportional to $t\mathcal{J}_0(K_0)/\omega$ results from the rotation (\ref{R2}). It reproduces the Schrieffer-Wolff transformation matrix for $K_0=0$ and $\omega=U$
\begin{equation}
\left.K_{\mathrm{rot}}^{(1)}(\tau)\right|_{K_0=0,\omega=U}=-i\frac{t}{U}\left(e^{iU\tau}h_+ - e^{-iU\tau}h_-\right)
\end{equation}
 where the matrices $h_{\pm}$ are given by
\begin{equation}
h_+=\begin{pmatrix} 0 & -1 & 1 & 0 \\
                  0 & 0 & 0 & 0 \\
									0  & 0 & 0 & 0 \\
									0 & -1 & 1 & 0 \end{pmatrix},\ 
h_-=\begin{pmatrix} 0 & 0 & 0 & 0 \\
                  -1 & 0 & 0 & -1 \\
									1  & 0 & 0 & 1 \\
									0 & 0 & 0 & 0 \end{pmatrix}
\end{equation}

They describe hopping processes between the Mott bands where the double occupancy is increased or reduced by one, respectively. 

We have verified that the next order in the expansion of the kick operator contains terms which scale like $t^2/\omega^2$ and $t(\omega-U)/\omega^2$.

\renewcommand{\theequation}{B\arabic{equation}}
\makeatletter
\renewcommand{\thefigure}{B\@arabic\c@figure} 
\renewcommand{\thetable}{B\@arabic\c@table} 
\setcounter{figure}{0}  

\section{Numerical simulation}
\label{ap:numerical}

\begin{figure}[t]%
\includegraphics{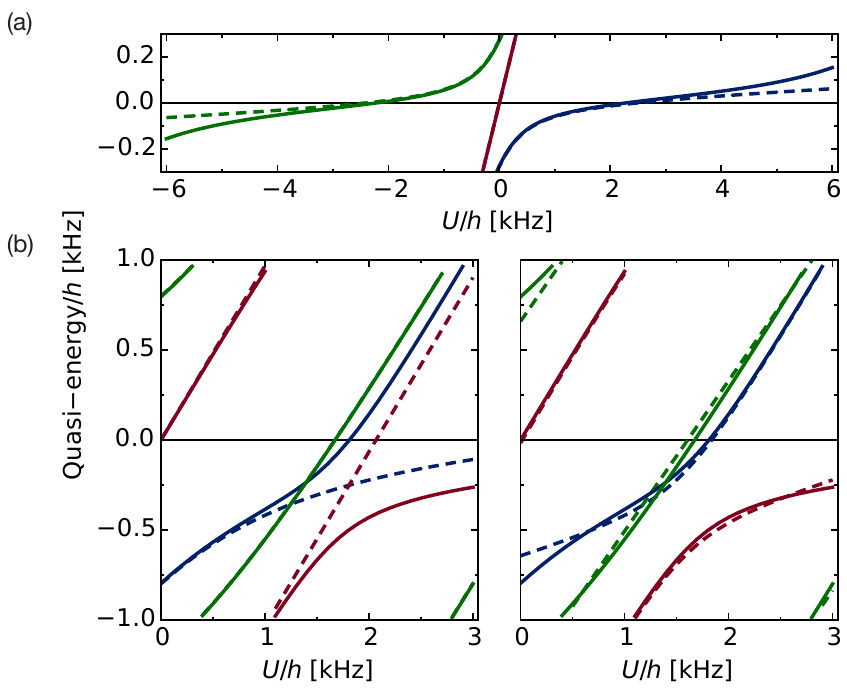}%
\caption{Quasi-energy spectrum, calculated analytically (dashed line) or numerically (full line). As the \ket{\mathrm{t}} state is unaffected by the modulation, it is omitted for clarity. In (a), the frequency $\omega/2\pi=8000$ Hz is much larger than the tunneling $t/h=548$ Hz, and both methods agree well ($K_0=1.94$). In (b) the frequency $\omega/2\pi=2000$ Hz is lower and $K_0=0.60$, $t/h=445$. The off-resonant analytic derivation (left) does not apply anymore, and must be replaced by the near-resonant prediction (right).}%
\label{fig:ThVsNum_sp}%
\end{figure}

In addition to the analytic derivation of the effective Hamiltonian, we also performed a numerical simulation of the two-site Hubbard model.
In this way, the off-resonant and near-resonant regimes presented above can be studied simultaneously, and the micromotion of all observables can be obtained.
To perform this calculation, we used a Trotter decomposition to evaluate the evolution operator over one period $\hat{U}(T+\tau_0,\tau_0)$, which evolves a quantum state from an initial time $\tau_0$ to time $\tau_0+T$.
The time-dependent Hamiltonian $\hat{H}(\tau)$ is approximated by $\hat{H}(\tau_j)$, which is piece-wise constant on $N$ consecutive time intervals $[\tau_j,\tau_{j+1}[$, with $\tau_j = j\,T/N+\tau_0$ and $0\leq j < N$.
The evolution operator can then be written as (we set $\hbar = 1$)
\begin{equation}
\hat{U}(T+\tau_0,\tau_0) = e^{-i \hat{H}(\tau_{N-1})\,T/N}\times ... \times e^{-i \hat{H}(\tau_0)\,T/N}
\end{equation}
For the evaluation we chose typically $N=50$.
The eigenvalues $\lambda_v$ of this operator $\hat{U}(T+\tau_0,\tau_0)$ are directly related to the quasi-energies $\epsilon_v$ by 
\begin{equation}
\lambda_v = \exp(-i \epsilon_v T)
\end{equation}
see also Eqs.\:(\ref{Defv}) and (\ref{vLab}).
However, the eigenvectors \ket{v(\tau_0)} are not uniquely defined, and depend on the starting phase $\tau_0$ of the periodic drive. 
To fully describe the driven system, it is necessary to obtain as well the evolution of the quantum state during the modulation cycle, which is given by $\ket{v(\tau_j)} = \hat{U}(\tau_j,\tau_0)\ket{v(\tau_0)}$.
By construction, $\ket{v(\tau_N)} = \ket{v(T+\tau_0)} = \ket{v(\tau_0)}$.
With this time-dependent state, we can then evaluate the instantaneous expectation value of any observable $\hat{O}$ in a Floquet state  by 
\begin{equation}
\langle \hat{O} \rangle (\tau_j) = \bra{v(\tau_j)}\hat{O}\ket{v(\tau_j)}
\end{equation}
in analogy to Eq.\:(\ref{TimeDepObs}).
This contains the full information about the evolution of the observable, in particular the amplitude of the micromotion can be determined from the Fourier components at the modulation frequency and its multiples.

\begin{figure}[t]
\includegraphics{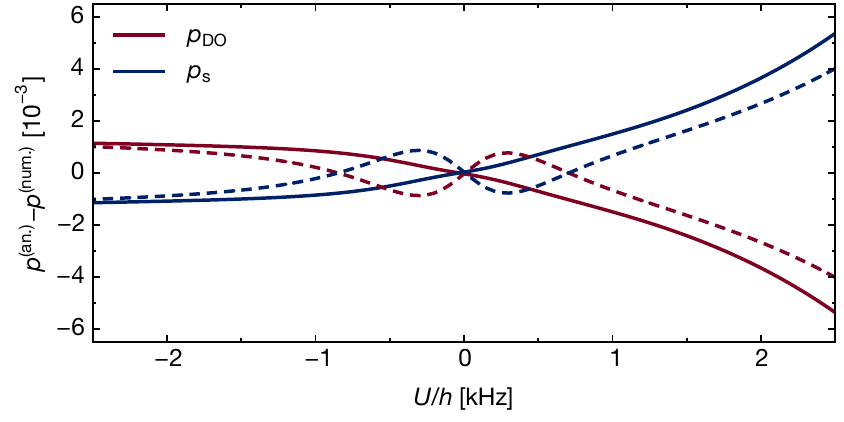}%
\caption{Off-resonant driving of the double well. We show the difference between the numerical and analytic predictions of \pdo (red) and \ps (blue) in the Floquet	 state originating from the static ground state (\ket{\tilde{\mathrm D}_+} for $U<0$, \ket{\tilde{\mathrm s}} for $U>0$), as a function of the interaction strength, for $t/h = 548$ Hz and $K_0 = 1.9$. The expansion up to first order in $1/\omega$ is shown in full line, and to order $1/\omega^2$ in dashed line. All calculations include an average over the micromotion, similar to the procedure used experimentally. For the whole range of interactions, the difference between the two predictions is smaller than 0.01, and begins to increase as $U$ approaches the resonance conditions $U \approx\omega$.}%
\label{fig:ThVsNum}%
\end{figure}

\subsection{Comparison to the analytic prediction}
As the numerical prediction is not limited to a certain range of interactions, we can compute the exact spectrum and expectation values of the observables and compare it to the analytic derivation presented above.

We first consider the off-resonant modulation for the experimental parameters used in Fig.\:\ref{fig1} of the main text. Here, the modulation frequency $\omega/2\pi = 8000$ Hz is much higher than the tunneling $t/h=548$ and the interaction $U/h$ is varied between $\pm 2800$ Hz. The quasi-energy spectrum obtained from the analytic derivation up to order $1/\omega^2$ (see Tables \ref{CorrOffRes} and \ref{CorrRes}) and the exact numerical eigenvalues are shown in Fig. \ref{fig:ThVsNum_sp}(a). Both methods agree quite well in the interaction range used in the experiment, while for $\left|U\right|>3000$ Hz, the analytic result starts to deviate from the exact result as the resonance at $U\approx \pm\omega$ is approached. Note that even for interactions as low as $U\approx 2250$ Hz, the singlet state becomes higher in energy than the triplet state, which is clearly beyond the scope of the lowest order effect that simply replaces $t\longrightarrow t\mathcal{J}_0(K_0)$.

Nevertheless, this deviation has a negligible effect on the expectation value of our observables taken in the Floquet state originating from the static ground state, which is shown in Fig. \ref{fig:ThVsNum}. Even if the effective Hamiltonian is approximated by the lowest order (\ref{LowOrderOffRes}), the double occupancy and singlet fraction differ by less than 0.01 from the exact result. However, due to the inversion of the lowest to energy levels, it is possible that singlets are converted into triplets if there are symmetry breaking terms present like a magnetic field gradient or a residual coupling to higher bands.

When the modulation frequency is decreased to $\omega/2\pi = 2000$ Hz, the analytic derivation in the off-resonant case differs significantly from the numerical evaluation for $U>1000$ Hz (see Fig. \ref{fig:ThVsNum_sp} left). Instead, the near-resonant description must be used, which agrees well with the numerical evaluation in the vicinity of the resonance $U\approx\omega$ (see Fig. \ref{fig:ThVsNum_sp} right).

\begin{figure}[t]%
\includegraphics{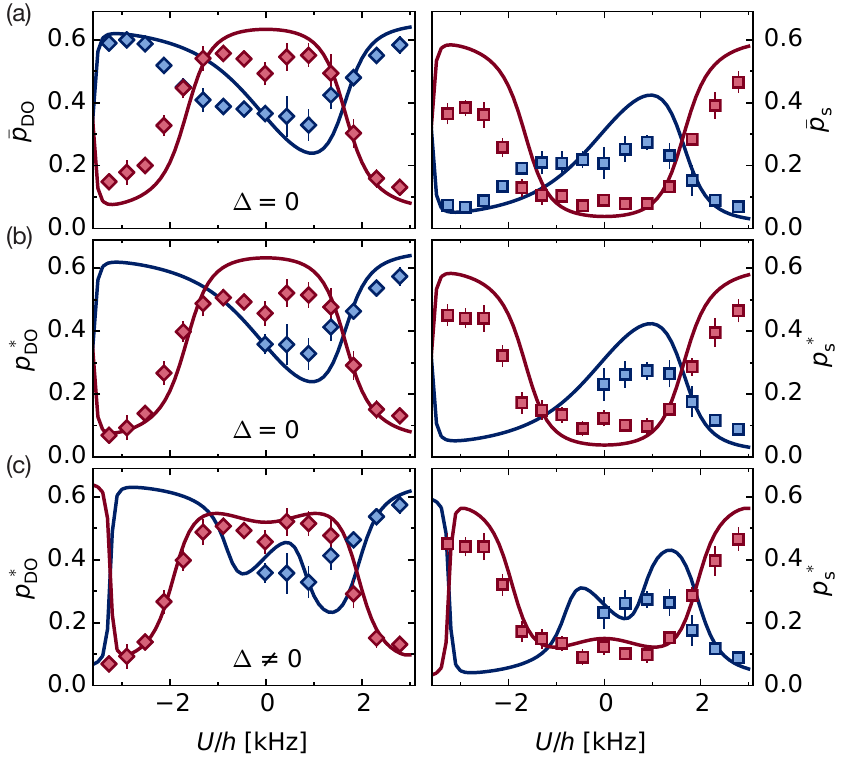}%
\caption{The experimental data shown in Fig. \ref{fig3b} in the main text (diamonds) is compared a numerical calculation of the system response. which includes an average over the micromotion (full lines), both for the double occupancies \pbdo (left) and for the singlet fraction \pbs (right). The red data corresponds to $\Uload/h=2800(80)$ Hz, and the blue data to $\Uload = 890(50)$ Hz.
The simulated response is rescaled to range between $0.018$ and $0.66$ \pbdo, and between $0.011$ and $0.64$ \pbs, to account for the starting conditions of the experiment. In panel (a), the raw experimental data is shown, and a balanced double well with $\Delta = 0$ is used for the calculation. In panel (b), the finite fidelity of the ramps is taken into account in the experimental data, and we show $p^{\ast} = \bar{p}-(\bar{p}^{(R)}-\bar{p}^{(R)}(\Uload))/2$ instead. Finally, in panel (c), the calculation is performed with an energy bias $\Delta/h = 800$ Hz between the double wells.} 
\label{fig:branchesNum}%
\end{figure}

\subsection{Comparison to the observed Floquet states}
The numerical results can also be confronted to the experimental results, for example as in the configuration of Fig. \ref{fig3b} of the main text, where an analytic prediction cannot be obtained for the full range of interactions.
We show in Fig. \ref{fig:branchesNum}(a) the experimental data for both $\Uload/h = 890$ Hz and $\Uload/h = 2800$ Hz, along with the corresponding numerical prediction. 
Part of the discrepancy between the two is due to the imperfect adiabaticity of the interaction ramps.
We assume that the excess of double occupancies and singlets observed in the return fractions \pbdor and \pbsr is generated uniformly during the interaction ramps (see Fig. \ref{fig3b}), and account for this imperfection by considering instead $p^{\ast} = \bar{p}-(\bar{p}^{(R)}-\bar{p}^{(R)}(\Uload))/2$, where $p$ can designate either \pdo or \ps.
The effect of this correction is shown in Fig. \ref{fig:branchesNum}(b).
Furthermore, the deviation between numerical prediction and experiment around $U=0$ can be qualitatively explained by the presence of a residual energy bias between the two wells.
We show in Fig. \ref{fig:branchesNum}(c) the numerical calculation with a potential bias $\Delta/h = 800$ Hz.
In our experiment, even though the mean potential bias may not be as large, the harmonic confinement introduces an inhomogeneous potential bias.
It can be modelled with an average bias given by $\bar{\Delta} = \int n(r) m \omega_{\mathrm{harm}}^2\, |r|\, d\, \mathrm{d}r/ \int n(r)\,\mathrm{d}r$, with $n(r)$ the probability that a double well located at distance $r$ from the origin is populated, and $\omega_{\mathrm{harm}}/2\pi = 114$ Hz the harmonic confinement frequency.
In our system, at zero temperature, $\bar{\Delta} = h\times 360$ Hz. This value can be increased both by the finite temperature of the sample and by a remaining potential bias \cite{si}.

\begin{figure}[t]
\includegraphics{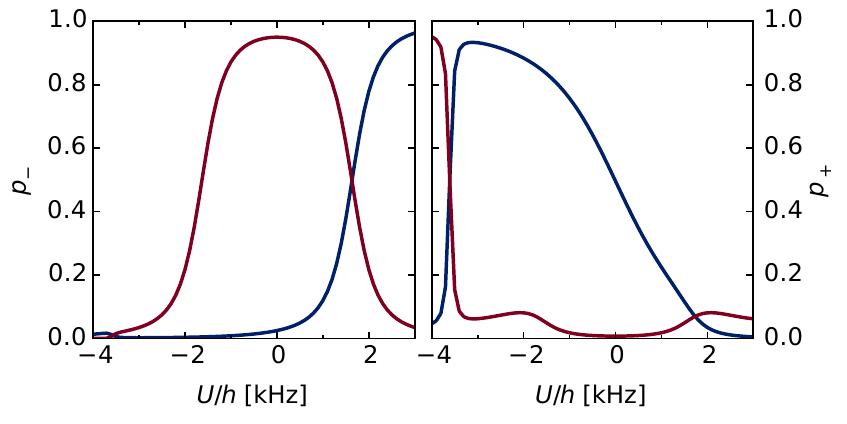}
\caption{With the numerical simulation of the system, we can also access the projection of the atomic state onto \ket{\mathrm{D}_-} (left) and \ket{\mathrm{D}_+} (right). The red line is the Floquet state accessed by setting $\Uload/h=2800$ Hz, and the blue line corresponds to $\Uload = 890$ Hz. In this parameter regime, the \ket{\mathrm{D}_+}state can only be accessed by choosing $\Uload/h = 890$ Hz, while the \ket{\mathrm{D}_-} state can be accessed with both Floquet states, at least for $U>0$.}
\label{fig:Do_pm}
\end{figure}

The numerical results also allow us to supplement the experimental results by distinguishing the states \ket{\mathrm{D}_-} and \ket{\mathrm{D}_+}, as shown in Fig. \ref{fig:Do_pm}.
When the drive is ramped up at $\Uload = 2800$ Hz followed by an interaction ramp to the attractive side, the initial \ket{\mathrm{s}} state is first transferred to the \ket{\mathrm{D}_-} state when $U=\omega$, and then back to the \ket{s} state when $U=-\omega$.
Similarly, when the drive is ramped up at $\Uload = 890$ Hz, the initial \ket{s} state is transferred to the \ket{\mathrm{D}_-} state as interactions reach $U=\omega$, and to the \ket{\mathrm{D}_+} state when they are decreased and become attractive.

\renewcommand{\theequation}{S\arabic{equation}}
\makeatletter
\renewcommand{\thefigure}{S\@arabic\c@figure} 
\renewcommand{\thetable}{S\@arabic\c@table} 
\setcounter{figure}{0}  
\setcounter{subsection}{0}

\appendix
\section*{Supplemental material}

\subsection{General preparation and optical lattice}

For our experiments we initially create a balanced two-component spin mixture of $^{40}$K fermions in the two magnetic sublevels $m_F = -9/2, -7/2$ of the $F = 9/2$ hyperfine manifold, which is confined in a harmonic optical dipole trap. 
We evaporatively cool the mixture to a quantum degenerate cloud with repulsive interactions of $115.6(8)\,a_0$ ($a_0$ denotes the Bohr radius) consisting of an atom number of $159(10)\times 10^3$ (15\% systematic error) at a temperature of 0.06(1)\:$T/T_{\mathrm{F}}$ ($T_{\mathrm{F}}$ denotes the Fermi temperature).
We can tune the scattering length with the Feshbach resonance located at 202.1 G.

The three-dimensional optical lattice is created by a combination of retro-reflected interfering and non-interfering laser beams of wavelength $\lambda=1064\,\mathrm{nm}$ and is described by the following potential \cite{Tarruell2012}:  
\begin{eqnarray} V(x,y,z) & = & -V_{\overline{X}}\cos^2(k
x+\theta/2)-V_{X} \cos^2(k
x)\nonumber\\
&&-V_{\widetilde{Y}} \cos^2(k y) -V_{Z} \cos^2(k z) \nonumber\\
&&-2\alpha \sqrt{V_{X}V_{Z}}\cos(k x)\cos(kz)\cos\varphi , 
\label{Lattice}
\end{eqnarray}
with $k=2\pi/\lambda$.
The lattice depths of each single beam in direction $x,y,z$ are given by $V_{\overline{X},X,\widetilde{Y},Z}$ in units of the recoil energy $E_R=h^2/2m\lambda^2$ ($h$ is the Planck constant and $m$ the mass of the atoms).
The visibility $\alpha=0.92(1)$ is measured via amplitude modulation spectroscopy with a $^{87}\mathrm{Rb}$ Bose-Einstein condensate in different interfering lattice configurations.  
We regulate the lattice potential such that $\theta=\pi\times 1.000(2)$. 
To fix the geometry of the lattice, the relative phase $\varphi$ of the two orthogonal retro-reflected beams X and Z is actively stabilized to $\varphi=0.00(3)\pi$. 
The lattice depths $V_{\overline{X},X,\widetilde{Y},Z}$ are independently calibrated using Raman–Nath diffraction on a $^{87}\mathrm{Rb}$ Bose–Einstein condensate.

\subsection{Preparation of the ground state in an array of double wells}

In the following we describe the preparation scheme for the ground state in the array of double wells that consists of different loading and formation steps.
Before loading the fermions into the optical lattice we tune the interactions to a large attractive value of $-3000(600)\,a_0$. 
We use an S-shaped lattice ramp of $200$ ms to load the atoms into the lowest band of a checkerboard configuration with lattice depths of $V_{\overline{X},X,\widetilde{Y},Z}=[0,3,7,3]E_R$ \cite{Tarruell2012}.
This is followed by a linear lattice ramp within $30$ ms to a $V_{\overline{X},X,\widetilde{Y},Z}=[0,30,30,30]E_R$ deep checkerboard 
lattice. 
Owing to the large attractive interactions, $68(3)$ \% of the atoms form double occupancies during this loading process.
For the splitting of the lattice sites we first tune the scattering length to either $-120(6)\,a_0$ or $105.5(9)\,a_0$ for measurements in the final lattice with attractive or repulsive interactions. 

Each lattice site is then subsequently split into a double well within $10$ ms by a linear ramp which increases $V_{\overline{X}}$ and decreases $V_{X}$ simultaneously, while the lattice depths in $y,z$-direction are kept constant.
The final lattice depths $V_{\overline{X},X, \widetilde{Y}, Z }$ and thus the tunnelling rate $t$ inside the dimer slightly vary and are given explicitly in Table \ref{table:lattices} for each measurement. 
The splitting process allows us to create an array of double wells with a tunnelling amplitude to neighbouring dimers below $h\times3$ Hz.
During this creation of dimers, the initially prepared double occupancies are smoothly transformed into the ground state of the double wells.
In a final step we ramp the on-site interactions in $5$ ms to the desired final value $U$ which allows us to prepare the lowest state of the static double well for all values of $U/t$ shown in Fig. 1(a).

\begin{table}%
\centering
\begin{tabular}{c|c|c}
Main text figure & 1 & 2,4,5 \\
\hline
$V_{\overline{X}}/E_R$ & 17.3 (5) & 17.6 (5) \\[2pt]

$V_X/E_R$ & 1.16 (3) & 0.96 (3) \\[2pt]

$V_{\widetilde{Y}}/E_R$ & 27.4 (8)  & 26.4 (8) \\[2pt]

$V_Z/E_R$  & 26.7 (9) & 28.4 (9) \\[2pt]

$t_{\mathrm{th}}/h$ (Hz) & 680 (100) & 490 (70) \\[2pt]

$t_{\mathrm{exp}}/h$ (Hz) & 550 (20)	& 450 (10)\\

$\Delta/h$ (Hz) & 0 (230) & 0 (200) \\

\end{tabular}
\caption{Lattice parameters used for the measurements of the main text. In this range of lattice depths, a systematic error on the potential can strongly influence the predicted dimer tunnelling $t_{\mathrm{th}}$. For this reason, we also give the measured tunnelling $t_{\mathrm{exp}}$, which we obtain from the measurement of \pdo as a function of $U$ in the static lattice. Error bars denote the standard error, systematic in the case of the lattice depths, and statistical in the case of $t_{\mathrm{exp}}$. Furthermore, the residual uncertainty on $\theta$ may lead to a potential bias between the two wells $\Delta$.}
\label{table:lattices}
\end{table}

\subsection{Periodic driving}

The mirror used for retro-reflecting the $X$ and $\overline{X}$ lattice beams is mounted on a piezo-electric actuator, which allow for a controlled phase shift of the reflected beam with respect to the incoming lattice beam.
To enter the driven regime, we linearly ramp up a sinusoidal modulation of the lattice position with the piezo-electric actuator, with frequency $\omega/2\pi$ along the direction of the dimers such that $V(x,y,z,\tau)\equiv V(x-A\cos(\omega\tau),y,z)$, and then maintain a fixed displacement amplitude $A$.
To maintain the phase relation $\varphi$ between the X and Z lattice beams during modulation, the phase of the respective incoming beams is modulated at the same frequency as the piezo-electric actuators using acousto-optical modulators, and is maintained to $\varphi=0.0(1)\pi$.
The modulation of the position of the lattice also leads to a residual modulation of the lattice depth of $\pm 2\,\%$, which in turn modifies the tunnelling amplitude by $\pm 10 \,\%$. 
In the co-moving frame, this corresponds to a modulation of the potential bias within a dimer $\Delta(\tau) = K_0\,\hbar\omega \sin(\omega \tau)$, where $K_0=m\omega A d/\hbar$  is the normalised drive amplitude, with $d$ the distance between the two sites of the dimer \cite{Eckardt2005, Lignier2007}.
In general, the distance between two sites of a simple optical lattice is given by $\lambda/2$. 
However, in our lattice configuration, the two sites of the double well are closer to each other.
To estimate $d$, we consider a cut of the lattice potential (\ref{Lattice}) for $y=z=0$ which can be written as (assuming that $\varphi=0$ and $\theta=\pi$)
\begin{equation}
V(x,0,0)=V_{\mathrm{S}}\cos^2(2k'x)-V_{\mathrm{L}}\cos^2(k'x)
\label{1Dcut}
\end{equation}
with $V_{\mathrm{S}}=V_{\overline{X}}-V_X$, $V_{\mathrm{L}}=4\alpha \sqrt{V_{X}V_{Z}}$ and $k'=k/2$. Now we make the approximation that the lattice potential is separable in the x- and z-directions and treat this as a one-dimensional problem. This should be a valid approximation since $V_Z$ is very large and the coupling in this direction is negligible.
For the one dimensional potential we first determine the Wannier functions located on the left and right sides of the double well, which are derived as the eigenstates of the band-projected position operator \cite{Uehlinger2013}. The distance $d$ is then evaluated as the difference between the eigenvalues of two neighboring Wannier states. For the lattice used in Fig. 1 in the main text, we find $d = 0.72(1) \times \lambda/2$ and for the configuration in Figs. 2, 4, 5 we calculate $d = 0.76(1) \times \lambda/2$. The uncertainty of these values follows from the systematic error on the lattice depths given in Table\:\ref{table:lattices}.

\subsection{Detection scheme}
To characterize the state on a double well, we follow the same procedures as in \cite{Jordens2008} and \cite{Greif2013a}.
Once the desired state has been prepared, we ramp up the lattice depth to $V_{\overline{X},X,\widetilde{Y},Z}=[30,0,30,30]E_R$ within $100\,\mathrm{\mu s}$, in order to freeze the evolution of the state.
This sudden ramp leaves the initial state unchanged.

To measure the singlet and triplet fractions \ps and \pt, we first remove double occupancies in the lattice by a series of Landau-Zener transfers, as they hinder the detection process.
We then apply a magnetic field gradient, which lifts the energy degeneracy for atoms with opposite spins on neighboring sites. 
This induces a coherent oscillation between the singlet and the triplet states. 
The singlet-triplet oscillations (STO) have a frequency of $\nu = \Delta_{\mathrm{STO}}/h$, where $\Delta_{\mathrm{STO}}$ is the energy splitting, and are only visible if the initial amount of singlets and triplets is different, as their respective time evolution is exactly out of phase. 
After a variable oscillation time we remove the magnetic field gradient and merge pairs of adjacent sites into a single site. 
We use a $10\,\mathrm{ms}$ linear ramp from the deep cubic lattice into a deep checkerboard lattice ($V_{\overline{X},X,\widetilde{Y},Z}=[0,30,30,30]\,E_R$), where tunnelling is still suppressed and the number of sites is divided by two. 
During merging, the \ket{\mathrm{s}} and \ket{\mathrm{t}} states are mapped onto different bands due to their distinct symmetry of the two-particle wave function. 
The (spatially symmetric) singlet state is mapped to two atoms in the lowest band of the final state, while the (spatially anti-symmetric) triplet state evolves into a final state with one atom in the first excited band and one atom in the lowest band. 
By adjusting the oscillation time to a maximum or minimum of the STO we can then detect the number of singlet and triplet states by measuring the number of atoms on doubly occupied sites with both atoms in the lowest band.

The fraction of atoms forming double occupancies can be measured either in the initial cubic lattice $V_{\overline{X},X,\widetilde{Y},Z}=[30,0,30,30]E_R$, which corresponds to \pdo in the main text, or after merging neighbouring sites by following the procedure described above, corresponding to \ps and \pt.
In either case, we perform an interaction-dependent radio-frequency transfer of the $m_F = −7/2$ spin state on doubly occupied sites to the previously unpopulated $m_F = −5/2$ spin state.
After this step, we ramp down the optical lattice potential and dipole trap within 20ms and apply a magnetic field gradient to separate the different $m_F$ states in a Stern-Gerlach measurement during ballistic expansion. 
Finally we take an absorption image and apply Gaussian fits to the density distribution of each spin component to determine the number of atoms in each spin state.

\subsection{Micromotion}

To measure the dependence of our observables on the timescale of the drive, we adopt the following procedure.
When we freeze the evolution of the quantum state by ramping up the lattice depth, the phase of the modulation is given by $\phi_M = \omega \tau_M +\phi_0$, where $\phi_0$ is the launching phase of the periodic drive, and $\tau_M$ the duration of the drive sequence.
To vary $\phi_M$, we keep $\tau_M$ fixed, and instead vary $\phi_0$.
As we are ramping up the periodic drive over many modulation cycles, we should in principle not be sensitive to the launching phase $\phi_0$ itself, but only to $\phi_M$.
To verify this, we varied simultaneously $\phi_0$ and $\tau_M$ while keeping $\phi_M$ fixed, and did not observe any change in our observables as expected.

In Fig. 5 (d) when scanning the interaction strength we observe an oscillation at the driving frequency $\omega$, that even becomes dominant for $U=0$.
This frequency component can be explained by a remaining finite site-offset $\Delta$ or a residual amplitude modulation of the lattice depth. 
In the first case, the points $\omega \tau_M+\phi =0$ and $\omega \tau_M+\phi = \pi$ of the modulation cycle depicted in Fig. 5 (a) are not equivalent, and an oscillation at $\omega/2\pi$ can be observed.
This behaviour is most pronounced when $U$ is close to $0$ and the imbalance between the wells is the largest energy scale.
In the second case, our modulation scheme can also introduce a residual amplitude modulation on the lattice depth. 
This in turn leads to a modulation of the tunneling amplitude $t$, which causes an oscillation of our observables at the drive frequency $\omega$.

\subsection{U calibration}

In our lattice configuration, the typical extension of the Wannier function can be comparable to the scattering length for the strongest interactions. 
Thus, the on-site interaction strength $U$ may differ from the calculated value, as was observed in a previous experiment \cite{Uehlinger2013}.
Therefore, we directly measure $U$ for different scattering lengths.
To this end, we prepare the atoms in the double wells at the desired interaction strength, and modulate the lattice depth $V_Z$ at a frequency $\Omega$.
This modulation can resonantly either create or destroy double occupancies (depending on the sign of the interactions) when $h\Omega$ matches the energy difference between two states.
In a double well, there are two possible transitions, either between \ket{\tilde{\mathrm{s}}} and \ket{\tilde{\mathrm{D}}_+} at $h\Omega = \sqrt{U^2+16\,t^2}$ or between \ket{\tilde{\mathrm{s}}} and \ket{\tilde{\mathrm{D}}_-} at $h\Omega = (\sqrt{U^2+16\,t^2}+U)/2$ (here again, we refer to the states by their majority component). 
In the absence of an energy bias between the two wells, only the first transition is allowed. 
However, the presence of the harmonic confinement in our experiment leads to a space-dependent energy bias, which restores the second transition.
By measuring the location of these resonances for various scattering lengths, we then determine $U(a)$ over the full range of scattering lengths.

\subsection{Amplitude dependence and adiabaticity for off-resonant modulation}

\begin{figure}[hb]
\includegraphics{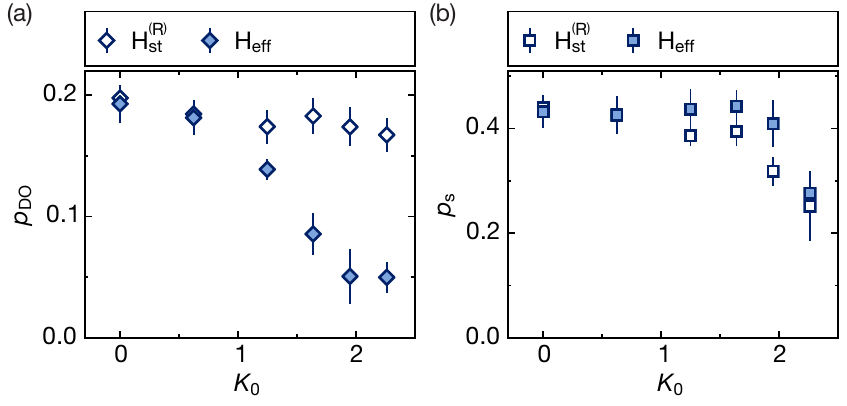}%
\caption{Amplitude dependence and adiabaticity for the off-resonant modulation (as in Fig. 1 (c,d) in the main text). The plot shows (a) the double-occupancy fraction $p_{\mathrm{DO}}$ (filled symbols) and the associated return fraction $p_{\mathrm{DO}}^{(R)}$ after reverting the modulation ramp (open symbols) and (b) the singlet fraction $p_{\mathrm{s}}$ (filled symbols) and return fraction $p_{\mathrm{s}}^{(R)}$ (open symbols) as a function of the shaking amplitude $K_0$ for a modulation frequency of $\omega/2\pi=8\:\mathrm{kHz}$ and $U/h = 1.5(1)\:\mathrm{kHz}$.}%
\label{fig:DOvsK0offRes}%
\end{figure}

Fig.\:\ref{fig:DOvsK0offRes} shows the dependence of the double-occupancy $p_{\mathrm{DO}}$ and singlet fraction $p_{\mathrm{s}}$ on the shaking amplitude $K_0$ for an off-resonant modulation frequency of $\omega/2\pi=8\:\mathrm{kHz}$ and $U/h=1.5(1)$ kHz. The double occupancy is decreasing as the tunneling is renormalized by the modulation, while the singlet fraction does not show a strong dependence up to $K_0\approx 2.0$ (compare Fig. 1\:(c,d) in the main text). The return fractions $p_{\mathrm{DO}}^{(R)}$ and $p_{\mathrm{s}}^{(R)}$ are comparable to the static observables apart from the regime of very strong driving with $K_0>2.0$. For this case, a significant loss of singlets and also atoms is observed, which we attribute to a residual coupling to higher bands.

\subsection{Singlet and triplet fraction and adiabatictiy for resonant modulation}

\begin{figure}[hb]%
\includegraphics{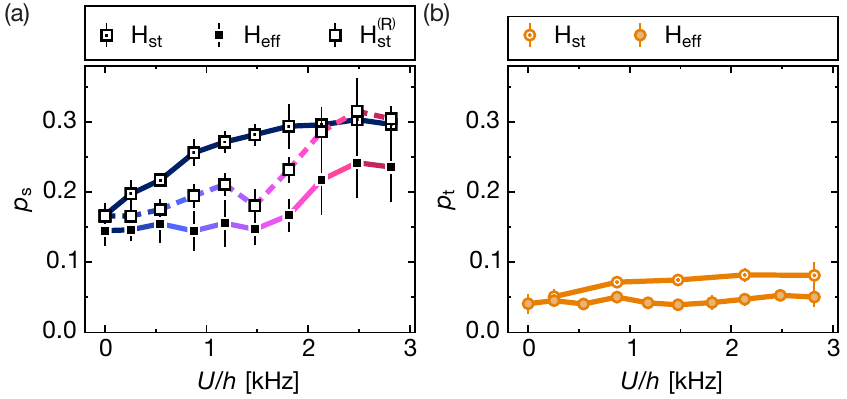}%
\caption{Resonance peak and adiabaticity for singlet and triplet fractions in the case of resonant modulation (as in Fig. 2 (a) in the main text). (a) Singlet and (b) triplet fractions for a shaking ramp time of $10\:\mathrm{ms}$ in the resonantly driven lattice with $\omega/2\pi=2\:\mathrm{kHz}$ and $K_0 = 1.14(2)$ (filled symbols), after reverting the loading ramp (open symbols) and in the static lattice (open-dotted symbols). Again, the deviation of the return fraction from the static values is indicative of a non-adiabatic process caused by the avoided crossing in the quasi-energy spectrum.}%
\label{fig:STOResonance}%
\end{figure}

As for the double occupancy, we also measure the singlet and triplet fractions $p_{\mathrm{s,t}}$ for the case of resonant modulation with a frequency of $\omega/2\pi=2\:\mathrm{kHz}$, which is depicted in Fig.\:\ref{fig:STOResonance}. Unlike the double occupancy (see Fig. 2(a) in the main text), the singlet fraction decreases compared to the static case when modulating with $\hbar\omega\approx U$, since we couple the singlet to a double occupancy state. The return fraction $\bar{p}_{\mathrm{s}}^{(R)}$ comes close to the static level only far away from the resonance, which confirms the observation that it is not possible to connect adiabatically to the Floquet states on the resonance by a simple ramp up of the modulation. The triplet fraction stays low for all interactions, both in the static and driven systems.

\end{document}